\documentclass[12pt,a4paper]{article}
\usepackage{graphicx}
\usepackage[T1]{fontenc}
\usepackage[utf8]{inputenc}
\usepackage{textcomp}
\usepackage[sc,osf]{mathpazo}
\usepackage{a4wide}  
\usepackage{latexsym,amsthm,amsfonts,amsmath,mathrsfs,amssymb}
\usepackage{dsfont}
\usepackage{accents}
\usepackage[nosort]{cite}
\usepackage{booktabs} % for much better looking tables
\usepackage[unicode,implicit]{hyperref}
\hypersetup{%
  pdftitle    = {Democratic actions with scalar fields: symmetric
    $\sigma$-models, supergravity actions and the effective theory of the
    type~IIB superstring}
  pdfkeywords = {duality, dual fields, electric-magnetic duality, pseudoaction,
  democratic action, type IIB, N=2B,d=10 supergravity, sigma models},
  pdfauthor   = {Jose Juan Fernandez-Melgarejo, Giacomo Giorgi, Carmen
    Gomez-Fayren, Tomas Ortin and Matteo Zatti},
  plainpages  = true,
  colorlinks  = true,
  citecolor   = blue,
  urlcolor    = red,
  linkcolor   = black
}
\newcommand{\hepth}[1]{{\tt
\href{http://www.arXiv.org/abs/hep-th/#1}{hep-th/#1}}}

\newcommand{\arxiv}[1]{{\tt arXiv:\href{http://www.arXiv.org/abs/#1}{#1}}}
\newcommand{\doi}[1]{{\tt DOI:\href{http://dx.doi.org/#1}{#1}}}

\allowdisplaybreaks

\makeatletter
\@addtoreset{equation}{section}
\makeatother

\begin{document}

\begin{flushright}
\small
IFT-UAM/CSIC-23-127\\
December 31\textsuperscript{st}, 2023\\
\normalsize
\end{flushright}

\vspace{.5cm}

\begin{center}

  {\Large {\bf Democratic actions with scalar fields: symmetric
    $\sigma$-models, supergravity actions and the effective theory of the
    type~IIB superstring}}

\vspace{1cm}

\renewcommand{\thefootnote}{\alph{footnote}}

{\sl\large Jos\'e J.~Fern\'andez-Melgarejo,$^{1,}$}\footnote{Email:
  {\tt melgarejo[at]um.es}}
{\sl\large Giacomo Giorgi,$^{1,}$}\footnote{Email:
  {\tt  giacomo.giorgi[at]um.es}}
{\sl\large Carmen G\'omez-Fayr\'en,$^{2,}$}\footnote{Email:
  {\tt carmen.gomez-fayren[at]estudiante.uam.es}}\\[.2cm]
{\sl\large Tom\'{a}s Ort\'{\i}n$^{2,}$}\footnote{Email:
  {\tt tomas.ortin[at]csic.es}}
{\sl\large and Matteo Zatti$^{2,}$}\footnote{Email:
  {\tt matteo.zatti[at]estudiante.uam.es}}

\setcounter{footnote}{0}
\renewcommand{\thefootnote}{\arabic{footnote}}
\vspace{1cm}

${}^{1}${\it Departamento de Electromagnetismo y Electr\'onica,\\
  Universidad de Murcia, Campus de Espinardo, 30100 Murcia, Spain}

\vspace{.4cm}

${}^{2}${\it Instituto de F\'{\i}sica Te\'orica UAM/CSIC\\
  C/ Nicol\'as Cabrera, 13--15,  C.U.~Cantoblanco, E-28049 Madrid, Spain}\\
\vspace{.8cm}

%%%%%%%%%%%%%%%%%%%%%%%%%%%%%%%%%%%%%%%%%%%%%%%%%%%%%%%%%%%%%%%%%%%%%%

{\bf Abstract}
\end{center}
\begin{quotation}
  {\small The dualization of the scalar fields of a theory into $(d-2)$-form
    potentials preserving all the global symmetries is one of the main
    problems in the construction of democratic pseudoactions containing
    simultaneously all the original fields and their duals. We study this
    problem starting with the simplest cases and we show how it can be solved
    for scalars parametrizing Riemannian symmetric $\sigma$-models as in
    maximal and half-maximal supergravities. Then, we use this result to write
    democratic pseudoactions for theories in which the scalars are
    non-minimally coupled to $(p+1)$-form potentials in any dimension. These
    results include a proposal of democratic pseudoaction for the generic
    bosonic sector of 4-dimensional maximal and half-maximal ungauged
    supergravities. Furthermore, we propose a democratic pseudoaction for the
    bosonic sector of $\mathcal{N}=2B,d=10$ supergravity (the effective action
    of the type~IIB superstring theory) containing two 0-, two 2-, one 4-, two
    6- and three 8-forms which is manifestly invariant under global
    SL$(2,\mathbb{R})$ transformations.}
\end{quotation}

\newpage
%%%%%%%%%%%%%%%%%%%%%%%%%%%%%%%%%%%%%%%%%%%%%%%%%%%%%%%%%%%%%%%%%%%%%%
%%%%%%%%%%%%%%%%%%%%%%%%%%%%%%%%%%%%%%%%%%%%%%%%%%%%%%%%%%%%%%%%%%%%%%
%%%%%%%%%%%%%%%%%%%%%%%%%%%%%%%%%%%%%%%%%%%%%%%%%%%%%%%%%%%%%%%%%%%%%%
%%%%%%%%%%%%%%%%%%%%%%%%%%%%%%%%%%%%%%%%%%%%%%%%%%%%%%%%%%%%%%%%%%%%%%
\pagestyle{plain}
%%%%%%%%%%%%%%%%%%%%%%%%%%%%%%%%%%%%%%%%%%%%%%%%%%%%%%%%%%%%%%%%%%%%%%
%%%%%%%%%%%%%%%%%%%%%%%%%%%%%%%%%%%%%%%%%%%%%%%%%%%%%%%%%%%%%%%%%%%%%%
%%%%%%%%%%%%%%%%%%%%%%%%%%%%%%%%%%%%%%%%%%%%%%%%%%%%%%%%%%%%%%%%%%%%%%
%%%%%%%%%%%%%%%%%%%%%%%%%%%%%%%%%%%%%%%%%%%%%%%%%%%%%%%%%%%%%%%%%%%%%%

\tableofcontents

%\newpage

%%%%%%%%%%%%%%%%%%%%%%%%%%%%%%%%%%%%%%%%%%%%%%%%%%%%%%%%%%%%%%%%%%%%%%
%%%%%%%%%%%%%%%%%%%%%%%%%%%%%%%%%%%%%%%%%%%%%%%%%%%%%%%%%%%%%%%%%%%%%%
%%%%%%%%%%%%%%%%%%%%%%%%%%%%%%%%%%%%%%%%%%%%%%%%%%%%%%%%%%%%%%%%%%%%%%
%%%%%%%%%%%%%%%%%%%%%%%%%%%%%%%%%%%%%%%%%%%%%%%%%%%%%%%%%%%%%%%%%%%%%%
\section*{Introduction}
%%%%%%%%%%%%%%%%%%%%%%%%%%%%%%%%%%%%%%%%%%%%%%%%%%%%%%%%%%%%%%%%%%%%%%
%%%%%%%%%%%%%%%%%%%%%%%%%%%%%%%%%%%%%%%%%%%%%%%%%%%%%%%%%%%%%%%%%%%%%%
%%%%%%%%%%%%%%%%%%%%%%%%%%%%%%%%%%%%%%%%%%%%%%%%%%%%%%%%%%%%%%%%%%%%%%
%%%%%%%%%%%%%%%%%%%%%%%%%%%%%%%%%%%%%%%%%%%%%%%%%%%%%%%%%%%%%%%%%%%%%%

$p$-branes naturally (\textit{electrically}) couple to $(p+1)$-form potentials
\cite{Green:1983wt,Green:1983sg,Bergshoeff:1987cm,Bergshoeff:1987qx,Polchinski:1995mt,Cederwall:1996pv,Cederwall:1996ri,Aganagic:1996pe,Aganagic:1996nn,Bergshoeff:1996tu,Aganagic:1997zq,Bandos:2000az}.\footnote{More
  references can be found in the reviews \cite{Duff:1994an,Ortin:2015hya}.}
However, the theories that describe the bulk dynamics of those $(p+1)$-form
potentials (supergravity theories, typically) are usually written in terms of
the lowest-rank potentials which are dual to them. Thus, $\mathcal{N}=2A,d=10$
supergravity, the effective fields theory of the type~IIA superstring, is
usually written in terms of the metric, dilaton (a scalar), NSNS 2-form, and
RR 1- and 3-forms plus a mass parameter (Romans') while the solitonic 5-brane
couples to the NSNS 6-form dual to the 2-form, and the D4-, D6- and D8-branes
couple to RR 5-, 7- and 9-forms dual, respectively, to the RR 3- and 1-forms
and to the mass parameter.  

Defining the higher-rank forms needed to describe the couplings of
higher-dimensional branes is always possible on-shell, providing equations of
motion for all of them.  It is always desirable to have an action from which
those equations of motion can be derived.\footnote{There is another reason why
  one may need the presence of the higher-rank forms in the action: in flux
  compactifications, their fluxes make relevant contributions
  \cite{Meessen:1998qm}. In particular, in
  Refs. \cite{Dibitetto:2019odu,Balaguer:2023jei} it is manifestly shown that,
  under the presence of D$p$/O$p$ systems, the modifications of the electric
  field strengths (and their Bianchi identities) induced by open string fluxes
  are read off from the couplings of the dual potentials to such objects.}
However, the field strengths of the higher-rank forms typically contain the
lower-rank ones and, very often, it is not possible to find an action for the
higher-rank forms because it must contain, at the same time, the lower-rank
ones, which are related to the former in a highly non-local way through the
duality relations.

The fact that an action for the higher-rank forms must also contain the
lower-rank ones can be turned into an advantage if one manages to give
consistence to the simultaneous presence of dual fields in the
action.\footnote{See Ref.~\cite{Avetisyan:2021heg} for the case of nonlinear
  electrodynamics and Ref.~\cite{Avetisyan:2022zza} for the extension to
  $(p+1)$-form potentials and their duals in arbitrary dimensions. The results
  of Ref.~\cite{Avetisyan:2022zza} could be used to formulate proper actions
  for the systems described in the current work via pseudoactions. For the
  free fields, this approach was introduced in Ref.~\cite{Mkrtchyan:2019opf}
  and discussed in detail for arbitrary $(p+1)$-form potentials and their
  duals in arbitrary dimensions in Ref.~\cite{Bansal:2021bis}.}
A solution to this problem is to use extensions of the Pasti-Sorokin-Tonin
formalism \cite{Pasti:1996vs} which is based on the introduction of an
auxiliary scalar field in the action. This method has been used in
Ref.~\cite{DallAgata:1997gnw,DallAgata:1998ahf}\footnote{See
  Ref.~\cite{Evnin:2022kqn} and, specially, the more recent
  Ref.~\cite{Ferko:2024zth} for a review} to construct covariant actions of
$\mathcal{N}=2B,d=10$ supergravity containing the 4-form but also 8-form duals
of the scalar fields and also in Refs.~\cite{Bandos:1997gd,Evnin:2023ypu} to
construct an action of $\mathcal{D}=1,d=11$ supergravity containing the 3-form
and its dual 6-form simultaneously. By dimensional reduction one can obtain an
action of $\mathcal{N}=2B,d=10$ supergravity containing the fundamental and
dual fields \cite{Bandos:2003et}.  In a slightly different context, it has
been used in Ref.~\cite{Bandos:1997ui} to construct a covariant worldvolume
action of the M5-brane.

An alternative solution, proposed in Ref.~\cite{Bergshoeff:2001pv}, consists
in including all the fields and treating them all on an equal footing as
independent (making it \textit{democratic}). This procedure doubles the
degrees of freedom and one has to impose by hand the twisted duality
\cite{Cremmer:1997ct,Cremmer:1998px} relations between
$(p+1)$- and $(d-p-3)$-forms only after the equations of motion have been
derived from the action. Since the duality relations are not derived from the
action, one is actually dealing with a \textit{pseudoaction}. The pseudoaction
introduced in Ref.~\cite{Bergshoeff:1995sq} for $\mathcal{N}=2B,d=10$
supergravity, which includes the 4-form with selfdual 5-form field strength
provides a good example: it contains an unconstrained (not selfdual) 4-form
which describes twice the degrees of freedom of the selfdual one and the
selfduality constraint must be imposed after the equations of motion have been
derived.

Each of these solutions\footnote{A more recent and different proposal can be
  found in Ref.~\cite{Mkrtchyan:2022xrm} for $\mathcal{N}=2A,B,d=10$
  supergravities.} presents advantages and disadvantages: the PST method
introduces unwanted auxiliary variables but gives a proper action from which
all the equations of motion can be derived while the second method does not
introduce unwanted auxiliary fields but only gives a pseudoaction. If one is
interested in evaluating the action on-shell (in order to study black-hole
thermodynamics, say), it is not clear whether the PST action gives the same
value as the original one. However, the democratic pseudoaction does, in
Euclidean signature, as we are going to discuss.

In this paper we are going to use the second method of dealing simultaneously
with fundamental and dual fields. Thus, our goal will be to construct
democratic pseudoactions containing all the fields and their duals whose
equations of motion give back the original ones upon use of duality
constraints.  Our main concern will be the dualization of the scalar fields,
which usually couple non-linearly among themselves and to other fields, into
$(d-2)$-form potentials.

The standard dualization procedure is only possible when the equation of
motion can be written, on-shell, as a total derivative. This happens to the
equation of motion of a given scalar field when there is a global symmetry of
the action acting on it, typically as a constant shift. The equation of
motion, then, is equivalent to the conservation of the associated Noether
current. Even if this is not immediately apparent, in that case the action can
be rewritten in terms of derivatives of the scalar and one can use the
Poincar\'e dualization method in the action. In absence of this kind of
symmetry, it is not known how to dualize the scalar field, but in supergravity
theories, there are typically many of these symmetries associated to
dualities.

When the theory contains several scalar fields which parametrize a non-linear
$\sigma$-model, things become more complicated. The shift symmetries are
isometries of the $\sigma$-model metric. One can always use coordinates
(scalar fields) adapted to a given isometry. In those coordinates the scalar
field shifted by the isometry does not occur explicitly in the metric and the
action can always be written in terms of its derivatives only. The equation of
motion will be a total derivative. One can only use coordinates adapted to
several isometries for those isometries that generate an Abelian
subgroup. However, even if the isometries do not commute, there are many
conserved currents as isometries and this guarantees that there are as many
combinations of the equations of motion as isometries that can be written as
total derivatives. These combinations can be used to define on-shell duals of
scalars.  Carrying it out this program in the full theory can be, in practice,
quite complicated. See, for example, Refs.~\cite{Yilmaz:2003as,Yilmaz:2003fp,Dereli:2005rn}.

Often (in all $d=4$ maximal and half-maximal supergravities, for instance),
the target space is a G/H Riemannian symmetric manifold, with more isometries
and conserved Noether currents than scalar fields. One can define a dual
$(d-2)$-form potential associated to each of the Noether currents (see, for
instance Ref.~\cite{DallAgata:1998ahf}), but, then, the number of dual fields,
dim$\, $G, would be larger than the number of original scalar fields,
dim$\, $G$-$dim$\, $H, which is not acceptable. However, since the set of all
dual $(d-2)$-forms transform in the adjoint representation of G, removing from
the action any number of them would break the global G-invariance of the
theory. This is one of the problems of the democratic pseudoaction of
$\mathcal{N}=2B,d=10$ supergravity proposed in Ref.~\cite{Bergshoeff:2001pv}:
only the RR scalar $C^{(0)}$ was dualized into the RR 8-form $C^{(8)}$ and,
therefore, the pseudoaction is not SL$(2,\mathbb{R})$-invariant as the
original theory.

A possible way out is to use a singular, but $G$-covariant, kinetic matrix in
the pseudoaction, as suggested in Ref.~\cite{Bandos:2016smv}. In this paper we
will identify the additional terms which are necessary to construct the
complete pseudoaction and we will use this result to construct
duality-invariant pseudoactions for several interesting theory, including all
the $d=4$ maximal and half-maximal supergravities and $\mathcal{N}=2B,d=10$
supergravity.

We are going to consider cases of increasing complexity: in
Section~\ref{sec-singlescalar} we start with the dualization of a single,
massless, real scalar $\phi$ coupled to gravity in $d$ spacetime dimensions,
to establish the notation and the basic facts. In Section~\ref{sec-sigmamodel}
we consider a generic non-linear $\sigma$-models with isometries and we will
study the dualization of scalars associated to an Abelian subgroup. This will
show us which are the needed additional terms mentioned above, which are the
first interesting results of this paper.  In
Section~\ref{sec-sigmamodelsymmetric}, we study the dualization of a
Riemannian symmetric $\sigma$-model and construct, using the additional terms
mentioned above and the singular but G-covariant kinetic matrix suggested in
Ref.~\cite{Bandos:2016smv}, the democratic pseudoaction that contains the
scalars that parametrize the G/H coset space and the dual $(d-2)$-form
potentials while preserving the global G invariance.  In
Section~\ref{sec-sigmamodelsymmetriccoupledtop+1forms} we apply this result to
the case in which the scalars are coupled to $(p+1)$-form potentials,
including the particular case $d=2(p+2)$, in which some of the transformations
in G are electric-magnetic dualities which leave invariant the equations of
motion but not the action. This particular case covers the bosonic sector of
all the maximal and half-maximal 4-dimensional supergravities. Finally, in
Section~\ref{sec-2B} we consider the case of $\mathcal{N}=2B,d=10$
supergravity, the effective field theory of the type~IIB superstring and
propose a pseudoaction that contains the dilaton and RR 0-form and a triplet
of 8-forms dual to them, the SL$(2,\mathbb{R})$ doublet of 2-forms (NSNS and
RR) and the dual doublet of 6-forms and a 4-form which is a SL$(2,\mathbb{R})$
singlet. The equations of all these fields derived from the pseudoaction
reduce to those of the fundamental fields when the (self-) duality constraints
are imposed on them.

Our conclusions and future directions of research are contained in
Section~\ref{sec-discussion}.

%%%%%%%%%%%%%%%%%%%%%%%%%%%%%%%%%%%%%%%%%%%%%%%%%%%%%%%%%%%%%%%%%%%%%%
%%%%%%%%%%%%%%%%%%%%%%%%%%%%%%%%%%%%%%%%%%%%%%%%%%%%%%%%%%%%%%%%%%%%%%
%%%%%%%%%%%%%%%%%%%%%%%%%%%%%%%%%%%%%%%%%%%%%%%%%%%%%%%%%%%%%%%%%%%%%%
%%%%%%%%%%%%%%%%%%%%%%%%%%%%%%%%%%%%%%%%%%%%%%%%%%%%%%%%%%%%%%%%%%%%%%
\section{Dualization of a single real scalar}
\label{sec-singlescalar}
%%%%%%%%%%%%%%%%%%%%%%%%%%%%%%%%%%%%%%%%%%%%%%%%%%%%%%%%%%%%%%%%%%%%%% 
%%%%%%%%%%%%%%%%%%%%%%%%%%%%%%%%%%%%%%%%%%%%%%%%%%%%%%%%%%%%%%%%%%%%%%
%%%%%%%%%%%%%%%%%%%%%%%%%%%%%%%%%%%%%%%%%%%%%%%%%%%%%%%%%%%%%%%%%%%%%%
%%%%%%%%%%%%%%%%%%%%%%%%%%%%%%%%%%%%%%%%%%%%%%%%%%%%%%%%%%%%%%%%%%%%%%

In order to establish the notation and describe what we want to do, it is
convenient to start with the simplest case, namely that of a single, massless,
real scalar, $\phi$, coupled to gravity, described by the Vielbein
$e^{a}=e^{a}{}_{\mu}dx^{\mu}$, in $d$ spacetime dimensions. The action that
dictates the dynamics of this system is\footnote{In this paper we are using
  differential-form language and the conventions of
  Ref.~\cite{Ortin:2015hya}.}

\begin{equation}
  \label{eq:singlescalaraction}
  S[e^{a},\phi] = \int\left\{
    (-1)^{d-1} \star (e^{a}\wedge e^{b}) \wedge R_{ab}
    +\tfrac{1}{2}d\phi\wedge \star d\phi\right\}\,.  
\end{equation}

\noindent
In this action $\star$ denotes the Hodge dual and, therefore,

\begin{equation}
  \star (e^{a}\wedge e^{b})
  =
  \frac{1}{(d-2)!}\epsilon_{c_{1}\cdots c_{d-2}}{}^{ab}
  e^{c_{1}}\wedge \cdots \wedge e^{c_{d-2}}\,.  
\end{equation}

\noindent
$\omega^{ab}=\omega_{\mu}{}^{ab}dx^{\mu}$ is the torsionless,
metric-compatible, Levi-Civita spin connection\footnote{It is antisymmetric
  $\omega^{ab}=-\omega^{ba}$ and satisfies
  $De^{a}= de^{a}-\omega^{a}{}_{b}\wedge e^{b}=0$.} and
$R^{ab}= \tfrac{1}{2}R_{\mu\nu}{}^{ab}dx^{\mu}\wedge dx^{\nu}$ is its
curvature 2-form

\begin{equation}
  \label{eq:curvaturedefined}
  R^{ab}
  \equiv
  d\omega^{ab} -\omega^{a}{}_{c}\wedge \omega^{cb}\,.
\end{equation}

The equations of motion which follow from this action are

\begin{subequations}
\begin{align}
        \label{eq:Easinglescalar}
  \mathbf{E}_{a}
  & =
    \imath_{a}\star (e^{c}\wedge e^{d})\wedge R_{cd}
      +\tfrac{(-1)^{d}}{2}
    \left(\imath_{a}d\phi\wedge \star d\phi
    +d\phi\wedge \imath_{a}\star d\phi\right)\,,
\\
& \nonumber \\
    \label{eq:E}
    \mathbf{E}
    & =
      -d\star d\phi\,.
\end{align}
\end{subequations}

Locally, the equation of motion of the scalar $\phi$ can be solved by
introducing a $(d-2)$-form $C$ such that

\begin{equation}
  \label{eq:singlescalardualityrelation}
G\equiv dC = \star d\phi\,.  
\end{equation}

\noindent
The equation of motion of the scalar $\phi$ becomes the Bianchi identity of
$G$ ($dG=0$) and the Bianchi identity of the scalar field strength $d\phi$
($d^{2}\phi=0$) becomes the equation of motion of the dual $(d-2)$-form $C$
($d\star G=0$).

Observe that the field strength $G$ is invariant under gauge transformations

\begin{equation}
  \label{eq:Sigmagaugetransformations}
\delta_{\Sigma}C = d\Sigma\,,  
\end{equation}

\noindent
where $\Sigma$ is an arbitrary $(d-3)$-form.

It is not difficult in this case to replace in the Einstein equations $d\phi$
by $\star G$\footnote{We have to take into
  account that, with our conventions, for a $(k+1)$-form $\omega^{(k+1)}$
  \begin{equation}
  \star^{2}\omega^{(k+1)} = (-1)^{k(d-1)}\omega^{(k+1)}\,,  
\end{equation}
and also that the canonical normalization of the action of a $k$-form with
$(k+1)$-form field strength $\omega^{(k+1)}$ is
\begin{equation}
\tfrac{(-1)^{dk}}{2} \omega^{(k+1)}\wedge \star \omega^{(k+1)}\,.
\end{equation}
} obtaining the equations of motion of a theory that contains the
metric and the $(d-2)$-form $C$ only:

\begin{subequations}
\begin{align}
        \label{eq:Easinglescalardual}
  \mathbf{E}_{a}
  & =
    \imath_{a}\star (e^{c}\wedge e^{d})\wedge R_{cd}
      +\tfrac{1}{2}
      \left(\imath_{a} G\wedge \star G +G\wedge \imath_{a}\star G\right)\,,
\\
& \nonumber \\
    \label{eq:Edual}
    \mathbf{E}
    & =
      -d\star G\,.
\end{align}
\end{subequations}

We can easily guess the action these equations of motion can be derived
from.\footnote{As usual in electric-magnetic duality, it is not possible to
  replace $\phi$ by its dual field $C$ directly in the action since the
  relation between these variables is non-local, even though the relation
  between their field strengths is. Substituting $d\phi$ by $\star dC$
  directly in the action leads to the wrong sign for the kinetic term of the
  dual field $C$.} However, there is a more systematic and direct procedure
(often called \textit{Poincar\'e duality}) that can be used as long as the
action depends only on the field strength $d\phi$ and not on the scalar field
$\phi$.\footnote{Sometimes it is possible to rewrite an action with explicit
  dependencies on $\phi$ in such a way that it only depends on $d\phi$. As a
  general rule, this happens when the action is invariant under constant
  shifts of $\phi$: $\phi\rightarrow \phi+c$. We will discuss this point in
  more detail later.} In these conditions, we can obtain an equivalent action
by replacing the scalar field $\phi$ by its 1-form field strength, which we
provisionally call $A$, as independent variable as long as we add a
Lagrange-multiplier term enforcing the Bianchi identity $dA=0$. This
constraint implies the local existence of $\phi$ and allows us to recover the
original scalar equation of motion. Calling $C$ this Lagrange multiplier and
defining $G\equiv dC$, the equivalent action takes the form

\begin{equation}
  S[e^{a},C,A] = \int\left\{
    (-1)^{d-1} \star (e^{a}\wedge e^{b}) \wedge R_{ab}
    +\tfrac{(-1)^{d}}{2}A\wedge \star A + G\wedge A\right\}\,.  
\end{equation}

The equation of motion of $A$ is algebraic:

\begin{equation}
 A =\star G\,,
\end{equation}

\noindent
and its solution can be used in the above action to get

\begin{equation}
  \label{eq:singlescalaractiondual}
  S[e^{a},C] = \int\left\{
    (-1)^{d-1} \star (e^{a}\wedge e^{b}) \wedge R_{ab}
    +\tfrac{(-1)^{d}}{2}G\wedge \star G\right\}\,,
\end{equation}

\noindent
which is the action from which the equations of motion
(\ref{eq:Easinglescalardual}) and (\ref{eq:Edual}) can be derived.

This action is invariant under the gauge transformations
Eq.~(\ref{eq:Sigmagaugetransformations}). Notice that, by following this
procedure, we have obtained the right sign for the kinetic term of $C$.

This dual action is not our real goal, though.  We are interested in actions
in which the original (``electric'') and the dual (``magnetic'') variables
appear simultaneously. Since this implies a redundancy of degrees of freedom,
it is necessary to use the relation between these variables
(Eq.~(\ref{eq:singlescalardualityrelation}), in this case) after the equations
of motion are derived from the action. Actions which need to be supplemented
by constraints in order to derive the equations of motion were called
\textit{pseudoactions} in Ref.~\cite{Bergshoeff:1995sq}. Thus, we are
interested in pseudoactions which contain both electric and magnetic variables
and which give equations of motion equivalent to those of the original theory
after the duality relations have been imposed. In the context of
$\mathcal{N}=2,A,B,d=10$ supergravity (the effective field theories of the
type~IIA and IIB superstrings), this kind of formulations of the theories were
called \textit{democratic} in Ref.~\cite{Bergshoeff:2001pv}. Thus, we are
interested in the democratic formulation of the theory given by the original
action Eq.~(\ref{eq:singlescalaraction}), which will be described by a
pseudoaction.

In this simple case, it is not difficult to see that the pseudoaction we are
after, containing $\phi$ and $C$ simultaneously, can be obtained by combining
the kinetic terms of $\phi$ and $C$ multiplied by $1/2$,\footnote{In this
  case, other, less symmetric combinations of coefficients of the kinetic
  terms of $\phi$ and $C$ give the same result.}

\begin{equation}
  \label{eq:singlescalaractiondemocratic}
  S[e^{a},\phi,C] = \int\left\{
    (-1)^{d-1} \star (e^{a}\wedge e^{b}) \wedge R_{ab}
    +\tfrac{1}{4}d\phi\wedge \star d\phi
    +\tfrac{(-1)^{d}}{4}G\wedge \star G\right\}\,,
\end{equation}

\noindent
and that it has to be supplemented by the constraint
Eq.~(\ref{eq:singlescalardualityrelation}). Indeed, if we use the duality
constraint in the equations of motion

\begin{subequations}
\begin{align}
  \mathbf{E}_{a}
  & =
    \imath_{a}\star (e^{c}\wedge e^{d})\wedge R_{cd}
          +\tfrac{(-1)^{d}}{4}
    \left(\imath_{a}d\phi\wedge \star d\phi
    +d\phi\wedge \imath_{a}\star d\phi\right)
    \nonumber \\
  & \nonumber \\
  & \hspace{.5cm}
      +\tfrac{1}{4}
      \left(\imath_{a} G\wedge \star G +G\wedge \imath_{a}\star G\right)\,,
\\
& \nonumber \\
    \mathbf{E}_{\phi}
    & =
      -\tfrac{1}{2}d\star d\phi\,,
\\
& \nonumber \\
    \mathbf{E}_{C}
    & =
      -\tfrac{1}{2}d\star G\,,
\end{align}
\end{subequations}

\noindent
to eliminate $C$, the energy-momentum $(d-1)$-form of $C$ becomes equal to
that of $\phi$ and one recovers the Einstein equation
(\ref{eq:Easinglescalar}) with the right coefficient and the equation of
motion of $C$ equation is automatically solved and one is left with the scalar
equation of motion with an overall factor of $1/2$. If, instead, one uses the
constraint to eliminate $\phi$ one obtains the dual result, \textit{i.e.}~the
equations of motion (\ref{eq:Easinglescalardual}) and (\ref{eq:Edual}), the
later with an overall factor of $1/2$.

Before ending this section, observe that using the duality relation directly
in the democratic pseudoaction does not lead to the original action because
the two kinetic terms simply cancel each other.\footnote{With different
  coefficients they may not cancel completely, but they will never give the
  original action back.} This implies that, if we try to evaluate the action
on-shell, since any solution satisfies the duality constraint, the
contributions of the kinetic terms will also cancel each other. However,
usually, it is the Euclidean action that one is interested in evaluating, not
the Lorentzian one. A field and its dual must necessarily have opposite
parities and one of them will by multiplied by $i$ when Wick-rotated, its
kinetic term acquiring an additional minus sign that will transform the
cancellation of the contributions of the dual kinetic terms into its addition.

%%%%%%%%%%%%%%%%%%%%%%%%%%%%%%%%%%%%%%%%%%%%%%%%%%%%%%%%%%%%%%%%%%%%%%
%%%%%%%%%%%%%%%%%%%%%%%%%%%%%%%%%%%%%%%%%%%%%%%%%%%%%%%%%%%%%%%%%%%%%%
%%%%%%%%%%%%%%%%%%%%%%%%%%%%%%%%%%%%%%%%%%%%%%%%%%%%%%%%%%%%%%%%%%%%%%
%%%%%%%%%%%%%%%%%%%%%%%%%%%%%%%%%%%%%%%%%%%%%%%%%%%%%%%%%%%%%%%%%%%%%%
\section{Non-linear $\sigma$-models with isometries}
\label{sec-sigmamodel}
%%%%%%%%%%%%%%%%%%%%%%%%%%%%%%%%%%%%%%%%%%%%%%%%%%%%%%%%%%%%%%%%%%%%%% 
%%%%%%%%%%%%%%%%%%%%%%%%%%%%%%%%%%%%%%%%%%%%%%%%%%%%%%%%%%%%%%%%%%%%%%
%%%%%%%%%%%%%%%%%%%%%%%%%%%%%%%%%%%%%%%%%%%%%%%%%%%%%%%%%%%%%%%%%%%%%%
%%%%%%%%%%%%%%%%%%%%%%%%%%%%%%%%%%%%%%%%%%%%%%%%%%%%%%%%%%%%%%%%%%%%%%

The basic dualization procedure used in the previous section will fail when
the action cannot be rewritten written in terms of the scalar field strength
only. This happens, generically, when the scalar field interacts with other
fields. We will consider the coupling of scalars to $(p+1)$-form potentials in
Section~\ref{sec-sigmamodelsymmetriccoupledtop+1forms} and now we will
consider interactions between several scalar fields.

When we have several scalar fields $\phi^{x}$ in our theory (in absence of
scalar potential), the situation becomes more complicated since the scalars
can couple non-trivially to the kinetic terms of other scalars. A convenient
way to describe all these possibilities in a geometric way is through the
non-linear $\sigma$-model formalism in which the scalar fields are interpreted
as mappings from spacetime to some ``target space'' in which they play the
role of coordinates. The couplings between scalars and kinetic terms are
collected in the $\sigma$-model (or target-space) metric $g_{xy}(\phi)$.  The
kinetic term (a combination of the kinetic terms of all the scalars and their
couplings)

\begin{equation}
\tfrac{1}{2}g_{xy}(\phi)d\phi^{x}\wedge \star d\phi^{y}\,,
\end{equation}

\noindent
can then be understood as the pull-back of the line element from the target
space to spacetime. Scalar field redefinitions can be reinterpreted as general
coordinate transformations in the target space.

The action of this system coupled to gravity takes the form

\begin{equation}
  \label{eq:sigmamodelaction}
  S[e^{a},\phi^{x}]
  =
  \int
  \left\{
    (-1)^{d-1} \star (e^{a}\wedge e^{b}) \wedge R_{ab}
  +\tfrac{1}{2}g_{xy}(\phi)d\phi^{x}\wedge \star d\phi^{y}\right\}\,,  
\end{equation}

\noindent
and the equations of motion are

\begin{subequations}
  \label{eq:EOMsigmamodelwithisometries}
  \begin{align}
        \label{eq:Easigmamodelwithisometries}
  \mathbf{E}_{a}
  & =
    \imath_{a}\star (e^{c}\wedge e^{d})\wedge R_{cd}
      +\tfrac{(-1)^{d}}{2}g_{xy}
    \left(\imath_{a}d\phi^{x}\wedge \star d\phi^{y}
    +d\phi^{x}\wedge \imath_{a}\star d\phi^{y}\right)\,,
\\
& \nonumber \\
    \label{eq:Exsigmamodelwithisometries}
    \mathbf{E}_{x}
    & =
      -g_{xy}\left[d\star d\phi^{y}
      +\Gamma_{zw}{}^{y}d\phi^{z}\wedge \star d\phi^{w}\right]\,,
  \end{align}
\end{subequations}

\noindent
where $\Gamma_{zw}{}^{y}$ are the components of the Christoffel symbols of the
target-space metric $g_{xy}$. The scalar equation of motion is, then, the
pullback of the geodesic equation in target space.

We would like to dualize the scalars $\phi^{x}$ into $(d-2)$-form fields. In
the single scalar case, we used the fact that the equation of motion
$d\star d\phi=0$ could be understood as a statement on the closedness of
certain differential form, $\star d\phi$ that we could locally solve by saying
that the differential form is exact $\star d\phi=dC$. In this theory, though,
the equations of motion of the scalars only have that form if the Christoffel
symbols take a very particular form. Since they are not tensors, this depends
very strongly on the coordinates (scalar fields, $\phi^{x}$) chosen, which
complicates the problem of finding out which scalars can be dualized and when.

There is, however, a coordinate-invariant characterization of the scalars that
can be dualized based on the following observation: scalar equations of motion
which are equivalent to the closedness of a $(d-1)$-form can be interpreted as
the conservation law of a $(d-1)$-form current $J$,

\begin{equation}
dJ=0\,.
\end{equation}

\noindent
If the theory is invariant under global symmetries acting on the scalar
fields, Noether's theorem ensures that there will be as many conserved
currents as symmetries. The on-shell conservation laws of these currents will
be combinations of some of the equations of motion of the scalar fields that
can be used to dualize them.

In order to characterize the scalar symmetries of the theory we denote by

\begin{equation}
  \label{eq:transformationscalars}
\delta_{A}\phi^{x} = k_{A}{}^{x}(\phi)\,,  
\end{equation}

\noindent
their infinitesimal generators. The indices $A$ label the independent
symmetries\footnote{That is: they take values in the adjoint representation of
  the Lie algebra of the symmetry group $G$.} and $k_{A}{}^{x}(\phi)$ are some
given (not arbitrary) functions of the scalar fields, as the symmetries we are
considering are global (not local) and take the form

\begin{equation}
\delta \phi^{x} =\alpha^{A}k_{A}{}^{x}(\phi)\,,
\end{equation}

\noindent
for constant, infinitesimal parameters $\alpha^{A}$.

It is not hard to see that the action Eq.~(\ref{eq:sigmamodelaction}) is
invariant under the transformations Eq.~(\ref{eq:transformationscalars}) if
and only if the $k_{A}{}^{x}(\phi)$ are Killing vectors of the target-space
metric, \textit{i.e.}~if

\begin{equation}
  \label{eq:Killingvectorequation}
\nabla_{(x|}k_{A\,|y)}=0\,,  
\end{equation}

\noindent
where $\nabla_{x}$ is the target-space covariant derivative with the
connection $\Gamma_{xy}{}^{z}$ and $k_{A\,x}=k_{A}{}^{y}g_{yx}$. The
associated Noether current $(d-1)$-forms are given by

\begin{equation}
\label{eq:NoethercurrentsjA}
J_{A} = \star \hat{k}_{A}\,,  
\end{equation}

\noindent
where $\hat{k}_{A}$ is the pullback of the 1-forms dual to the Killing vectors

\begin{equation}
  \hat{k}_{A}
  \equiv
  k_{A}{}^{x}g_{xy}d\phi^{y}\,.  
\end{equation}

\noindent
Furthermore, it is not difficult to see, using the Killing equation, that

\begin{equation}
  \label{eq:eomdja}
k_{A}{}^{x}\mathbf{E}_{x} = -dJ_{A}\,,  
\end{equation}

\noindent
which establishes the relation between the scalar equations of motion and the
on-shell conservation of the Noether currents we were looking for.

The above conservation laws suggest that we may try to define dual
$(d-2)$-forms $C_{A}$ with field strengths $G_{A}$ associated the conserved
currents via

\begin{equation}
  \label{eq:GAstarjA}
G_{A}\equiv dC_{A} = J_{A}\,.  
\end{equation}

\noindent
Since the currents $J_{A}$ only transform under global $G$ transformations,
$G_{A}$ must be gauge invariant and the $(d-2)$-forms only transform under
gauge transformations

\begin{equation}
  \delta_{\Sigma}C_{A}
  =
  d\Sigma_{A}\,.
\end{equation}

The currents $J_{A}$ do not occur explicitly in the action and, therefore, it
is not clear how one can use the Poincar\'e duality procedure. When the
Killing vectors $k_{A}{}^{x}$ commute, though, it is possible to use
coordinates adapted to all the isometries.\footnote{We can also restrict
  ourselves to an Abelian subgroup of the isometry group of the target space
  metric.} In this adapted coordinate system the target-space metric $g_{xy}$
is independent of the scalars associated to the Killing vectors and they can
be Poincar\'e-dualized in the standard fashion because those scalars only
occur through their field strengths. We are going to consider this particular
case first.

%%%%%%%%%%%%%%%%%%%%%%%%%%%%%%%%%%%%%%%%%%%%%%%%%%%%%%%%%%%%%%%%%%%%%%
%%%%%%%%%%%%%%%%%%%%%%%%%%%%%%%%%%%%%%%%%%%%%%%%%%%%%%%%%%%%%%%%%%%%%%
%%%%%%%%%%%%%%%%%%%%%%%%%%%%%%%%%%%%%%%%%%%%%%%%%%%%%%%%%%%%%%%%%%%%%%
%%%%%%%%%%%%%%%%%%%%%%%%%%%%%%%%%%%%%%%%%%%%%%%%%%%%%%%%%%%%%%%%%%%%%%
\subsection{Non-linear $\sigma$-models with commuting isometries}
\label{sec-sigmamodelcommuting}
%%%%%%%%%%%%%%%%%%%%%%%%%%%%%%%%%%%%%%%%%%%%%%%%%%%%%%%%%%%%%%%%%%%%%% 
%%%%%%%%%%%%%%%%%%%%%%%%%%%%%%%%%%%%%%%%%%%%%%%%%%%%%%%%%%%%%%%%%%%%%%
%%%%%%%%%%%%%%%%%%%%%%%%%%%%%%%%%%%%%%%%%%%%%%%%%%%%%%%%%%%%%%%%%%%%%%
%%%%%%%%%%%%%%%%%%%%%%%%%%%%%%%%%%%%%%%%%%%%%%%%%%%%%%%%%%%%%%%%%%%%%%

In this case we can use the machinery and notation (hats for original fields)
of Kaluza-Klein dimensional reductions for the target space metric. We choose
coordinates adapted to all the commuting isometries (all the Killing vectors
$k_{A}{}^{x}$) to be considered splitting the coordinate indices into those
related to the isometries, $A$, and the rest, $m$, $\{x\}=\{A,m\}$ and using
the notation $\{\hat{\phi}^{x}\} =\{\varphi^{A},\phi^{m}\}$. In these
coordinates, the components of the Killing vectors are
$k_{A}{}^{x} =\delta_{A}{}^{x}$
(\textit{i.e.}~$k_{A}{}^{B} =\delta_{A}{}^{B}\,,\,\,\,k_{A}{}^{m} =
0$).

By definition, the target-space metric only depends on the scalar fields
$\phi^{m}$ and its components and those of its inverse can be written in the
form

\begin{equation}
  \begin{aligned}
  \left(\hat{g}_{xy}\right)
    & =
    \left(
      \begin{array}{lr}
        g_{AB} & g_{AC}A^{C}{}_{n} \\
               & \\
        g_{BC}A^{C}{}_{m}\,\,\,\, & g_{mn}+A^{A}{}_{m}A^{B}{}_{n}g_{AB} \\
      \end{array}
    \right)\,,
    \\
    & \\
      \left(\hat{g}^{xy}\right)
    & =
    \left(
      \begin{array}{lr}
        g^{AB}+A^{A}{}_{m}A^{B}{}_{n}g^{mn} & -A^{A}{}_{p}g^{pn} \\
               & \\
        -A^{B}{}_{p}g^{pm}\,\,\,\, & g^{mn} \\
      \end{array}
      \right)\,,
  \end{aligned}
\end{equation}

\noindent
where

\begin{equation}
  g^{AB}g_{BC} = \delta^{A}{}_{C}\,,
  \hspace{1.5cm}
  g^{mn}g_{np} = \delta^{m}{}_{p}\,.
\end{equation}

We stress that all the target-space fields $g_{AB},g_{mn}$ and $A^{A}{}_{m}$
are independent of the scalars $\varphi^{A}$.

In terms of these new variables (actually, combinations of scalar fields), the
action Eq.~(\ref{eq:sigmamodelaction}) takes the form

\begin{equation}
  \label{eq:sigmamodelactionsplit}
  S[e^{a},\varphi^{A},\phi^{m}] = \int \left\{
    (-1)^{d-1} \star (e^{a}\wedge e^{b}) \wedge R_{ab}
    +\tfrac{1}{2}g_{AB}\mathcal{D}\varphi^{A}\wedge \star \mathcal{D}\varphi^{B}
    +\tfrac{1}{2}g_{mn}d\phi^{m}\wedge \star d\phi^{n}
    \right\}\,,  
\end{equation}

\noindent
where we have defined

\begin{equation}
  \mathcal{D}\varphi^{A}
  \equiv
  d\varphi^{A}+A^{A}{}_{m}d\phi^{m}\,.
\end{equation}

The equations of motion take now the form

\begin{subequations}
  \begin{align}
  \mathbf{E}_{a}
  & =
    \imath_{a}\star (e^{c}\wedge e^{d})\wedge R_{cd}
      +\tfrac{(-1)^{d}}{2}g_{AB}
    \left(\imath_{a}\mathcal{D}\varphi^{A}\wedge \star \mathcal{D}\varphi^{B}
    +\mathcal{D}\varphi^{A}\wedge \imath_{a}\star \mathcal{D}\varphi^{B}\right)
    \nonumber \\
    & \nonumber \\
    & \hspace{.5cm}
    +\tfrac{(-1)^{d}}{2}g_{mn}
    \left(\imath_{a}d\phi^{m}\wedge \star d\phi^{n}
    +d\phi^{m}\wedge \imath_{a}\star d\phi^{n}\right)\,,
\\
& \nonumber \\
    \mathbf{E}_{A}
    & =
      -d\left[g_{AB}\star\mathcal{D}\varphi^{B}\right]\,,
    \\
    & \nonumber \\
\mathbf{E}_{m}
    & =
      -g_{mn}\left[d\star d\phi^{n}
      +\Gamma_{pq}{}^{n}d\phi^{p}\wedge \star d\phi^{q} \right]
      +\tfrac{1}{2}\frac{\partial g_{AB}}{\partial \phi^{m}}
      \mathcal{D}\varphi^{A}\wedge \star \mathcal{D}\varphi^{B}
      \nonumber \\
    & \nonumber \\
    & \hspace{.5cm}
    +g_{AB}F^{A}{}_{mn}d\phi^{n}\wedge \star \mathcal{D}\varphi^{B}
    +A^{A}{}_{m}\mathbf{E}_{A}\,.
  \end{align}
\end{subequations}

The equations of motion of the scalars $\varphi^{A}$ can be understood as the
expression of the conservation of the Noether currents associated to the
invariance of the action under the constant shifts generated by the Killing
vectors because the currents $J_{A}$ are given by

\begin{equation}
  J_{A}
  \equiv g_{AB}
  \star \mathcal{D}\varphi^{B}\,,
  \,\,\,\,\,\,
  \Rightarrow
  \,\,\,\,\,\,
\mathbf{E}_{A}
=
-dJ_{A}\,.
\end{equation}

\noindent
Then, we can solve locally the equations of motion of those scalars by
introducing the dual $(d-2)$-forms $C_{A}$:

\begin{equation}
  \label{eq:dualityrelation}
G_{A} \equiv dC_{A} =  J_{A}\,.
\end{equation}

\noindent
The field strengths $G_{A}$ are invariant under gauge transformations of the
form

\begin{equation}
\label{eq:SigmaAgaugetransformations}
  \delta_{\Sigma}C_{A}=d\Sigma_{A}\,,
\end{equation}

\noindent
where the $\Sigma_{A}$ are $(d-3)$-forms.

The duality relation Eq.~(\ref{eq:dualityrelation}) together with the
definition of the currents $J_{A}$ can be used to express the 
field strengths of the scalars  $\varphi^{A}$ in terms of the dual
$(d-2)$-forms $C_{A}$ (and the rest of the scalars)

\begin{equation}
  \label{eq:dvarphiA}
d\varphi^{A} = g^{AB}\star G_{B} -A^{A}{}_{m}d\phi^{m}\,.    
\end{equation}

\noindent
Then, the Bianchi identity of these field strengths, $d^{2}\varphi^{A}=0$)
gives the equations of motion of $(d-2)$-forms $C_{A}$:

\begin{equation}
  d\left(g^{AB}\star G_{B}\right)  -F^{A}=0\,,  
\end{equation}

\noindent
where

\begin{equation}
  F^{A} \equiv \tfrac{1}{2}F^{A}{}_{mn}d\phi^{m}\wedge d\phi^{n}\,,
\hspace{1cm}
  \text{with}
\hspace{1cm}
  F^{A}{}_{mn}
  \equiv
  2\partial_{[m} A^{A}{}_{n]}\,.
\end{equation}

The scalar fields $\varphi^{A}$ can be completely eliminated from the action
by standard Poincar\'e dualization and the result is an action that contains
the field variables $C_{A}$ and $\phi^{m}$ (which we do not know how to
dualize) and which is invariant up to a total derivative under the
$\delta_{\Sigma}$ gauge transformations defined in
Eq.~(\ref{eq:SigmaAgaugetransformations})

\begin{equation}
  \label{eq:sigmamodelactiondual}
  \begin{aligned}
    S[e^{a},C_{A},\phi^{m}]
    & = \int \left\{ (-1)^{d-1} \star (e^{a}\wedge
      e^{b}) \wedge R_{ab} +\tfrac{(-1)^{d}}{2}g^{AB}G_{A}\wedge \star G_{B}
      +C_{A}\wedge F^{A}
    \right.
    \\
    & \\
    & \hspace{.5cm}
    \left.
      +\tfrac{1}{2}g_{mn}d\phi^{m}\wedge \star d\phi^{n}
    \right\}\,.
  \end{aligned}
\end{equation}

The equations of motion that follow from this
action are

\begin{subequations}
\begin{align}
  \mathbf{E}_{a}
  & =
    \imath_{a} \star (e^{c}\wedge e^{d}) \wedge R_{cd}
    +\tfrac{1}{2}g^{AB} \left(\imath_{a} G_{A} \wedge \star G_{B}
    +(-1)^{d} G_{A} \wedge \imath_{a} \star G_{B} \right)
  \nonumber \\
  & \nonumber \\
  & \hspace{.5cm}
    + \tfrac{(-1)^{d}}{2}g_{mn}(\imath_{a} d\phi^{m} \wedge \star d\phi^{n}
    +d\phi^{m} \wedge \imath_{a} \star d\phi^{n})\, ,
  \\
  & \nonumber \\
  \mathbf{E}^{A}
  & =
    -d(g^{AB}\star G_{B})+F^{A}\,,
  \\
  & \nonumber\\
  \mathbf{E}_{m}
  & =
    -g_{mn}[d\star d \phi^{n} +\Gamma_{pq}{}^{n} d\phi^{p} \wedge \star
    d\phi^{q}]
    +\tfrac{(-1)^{d}}{2} \frac{\partial g^{AB}}{\partial \phi^{m}}
    G_{A} \wedge \star G_{B}
      \nonumber \\
  & \nonumber \\
  & \hspace{.5cm}
    -2(-1)^{d-1}G_{A} \wedge F^{A}{}_{nm}d\phi^{n}\,, 
\end{align}
\end{subequations}

\noindent
are completely equivalent to those of the original fields upon use of the
duality relation Eq.~(\ref{eq:dualityrelation}).\footnote{The Bianchi identity
  of the target-space 2-form field strengths $F^{A}$ occur in the equation of
  motion of the $\phi^{m}$ and explains why the term $C_{A}\wedge F^{A}$ does
  not contribute to them.} Furthermore, we can construct a democratic action
by simply adding the original kinetic term of the $\varphi^{A}$s to this
action, changing the normalization of the kinetic terms to get the right
normalization of the energy-momentum tensor in the Einstein equation (the
topological term $C_{A}\wedge F^{A}$ does not contribute to it):

\begin{equation}
  \label{eq:sigmamodelactiondualdemocratic}
  \begin{aligned}
    S[e^{a},C_{A},\varphi^{S},\phi^{m}]
    & = \int \left\{ (-1)^{d-1} \star (e^{a}\wedge
      e^{b}) \wedge R_{ab}
      +\tfrac{1}{4}g_{AB}\mathcal{D}\varphi^{A}\wedge \star \mathcal{D}\varphi^{B}
    \right.
    \\
    & \\
    & \hspace{.5cm}
    \left.
      +\tfrac{(-1)^{d}}{4}g^{AB}G_{A}\wedge \star G_{B}
      +\tfrac{1}{2}C_{A}\wedge F^{A}
      +\tfrac{1}{2}g_{mn}d\phi^{m}\wedge \star d\phi^{n}
    \right\}\,.
  \end{aligned}
\end{equation}

The equations of motion that follow from this action are

\begin{subequations}
\begin{align}
  \mathbf{E}_{a}
  & =
    \imath_{a} \star (e^{c}\wedge e^{d}) \wedge R_{cd}
    +\tfrac{1}{4}g^{AB} \left(\imath_{a} G_{A} \wedge \star G_{B}
    +(-1)^{d} G_{A} \wedge \imath_{a} \star G_{B} \right)
    \nonumber \\
    & \nonumber \\
  & \hspace{.5cm}
    + \tfrac{(-1)^{d}}{2}g_{mn}(\imath_{a} d\phi^{m} \wedge \star d\phi^{n}
    +d\phi^{m} \wedge \imath_{a} \star d\phi^{n})
  \nonumber \\
  & \nonumber \\
  & \hspace{.5cm}
    +\tfrac{(-1)^{d}}{4}g_{AB}
    (\imath_{a} \mathcal{D}\varphi^{A}\wedge \star \mathcal{D}\varphi^{B}
    +\mathcal{D}\varphi^{A}\wedge \imath_{a} \star \mathcal{D}\varphi^{B})\,,
  \\
    &  \nonumber \\
  \mathbf{E}_{A}
  & =
    -\tfrac{1}{2}d(g_{AB}\star \mathcal{D}\varphi^{B})\,,
  \\
    &  \nonumber \\
  \mathbf{E}^{A}
  & =
    -\tfrac{1}{2}d(g^{AB}\star G_{B})+\tfrac{1}{2}F^{A}\,,
  \\
    &  \nonumber \\  
  \mathbf{E}_{m}
  & =
    -g_{mn}[d\star d \phi^{n} +\Gamma_{pq}{}^{n} d\phi^{p} \wedge \star d\phi^{q}]
    +\frac{1}{4}\frac{\partial g_{AB}}{\partial \phi^{m}}
    \mathcal{D}\varphi^{A} \wedge \star \mathcal{D} \varphi^{B}
    \nonumber \\
  & \nonumber \\
  & \hspace{.5cm}
    +\tfrac{(-1)^{d}}{4}\frac{\partial g^{AB}}{\partial \phi^{m}}
    G_{A} \wedge \star G_{B}
    +g_{AB}F^{A}{}_{mn}d\phi^{n} \wedge \star \mathcal{D}\varphi^{B}
    \nonumber \\
  & \nonumber \\
  & \hspace{.5cm}
-(-1)^{d-1}G_{A} \wedge F^{A}{}_{nm}d\phi^{n}\,.
   \end{align}
\end{subequations}

This is a simple case in which the scalars $\varphi^{A}$ can be completely
replaced by its dual $(d-2)$-forms $C_{A}$. Still, we have learned that it is
necessary to include the topological term $C_{A}\wedge F^{A}$ in the dual
action.

It is not clear at all how to dualize the rest of the scalars, if this is
possible at all. On general grounds we expect the scalars related to
symmetries to be ``dualizable'' because their equations of motion are related
to the conservation of certain Noether currents and then we can dualize on
shell those equations. If there are enough symmetries, we may be able to
dualize all the scalars, at least in the sense of being able to define the
$(d-2)$-form potentials dual to them.  However, when the isometries do not
commute, we cannot use coordinates adapted to all the isometries and we cannot
use the Poincar\'e dualization procedure and, any putative action containing
the dual fields should also include the original scalar fields. On the other
hand, if we do use all the currents (all the isometries) the dual $(d-2)$-form
potentials will not fill a linear representation of the symmetry group and the
invariance of the theory containing the dual fields under this group will in
general be broken.

We do not know how to solve this problem in general. However, in the case in
which the target space is a Riemannian symmetric manifold, inspired by the
form of the action that we have just constructed (the necessity of the
topological term $C_{A}\wedge F^{A}$), we have found a way to construct a
democratic action which manifestly preserves all the symmetries of the
original one. We describe this construction in the next section.

%%%%%%%%%%%%%%%%%%%%%%%%%%%%%%%%%%%%%%%%%%%%%%%%%%%%%%%%%%%%%%%%%%%%%%
%%%%%%%%%%%%%%%%%%%%%%%%%%%%%%%%%%%%%%%%%%%%%%%%%%%%%%%%%%%%%%%%%%%%%%
%%%%%%%%%%%%%%%%%%%%%%%%%%%%%%%%%%%%%%%%%%%%%%%%%%%%%%%%%%%%%%%%%%%%%%
%%%%%%%%%%%%%%%%%%%%%%%%%%%%%%%%%%%%%%%%%%%%%%%%%%%%%%%%%%%%%%%%%%%%%%
\section{Dualization of Riemannian symmetric $\sigma$-models}
\label{sec-sigmamodelsymmetric}
%%%%%%%%%%%%%%%%%%%%%%%%%%%%%%%%%%%%%%%%%%%%%%%%%%%%%%%%%%%%%%%%%%%%%% 
%%%%%%%%%%%%%%%%%%%%%%%%%%%%%%%%%%%%%%%%%%%%%%%%%%%%%%%%%%%%%%%%%%%%%%
%%%%%%%%%%%%%%%%%%%%%%%%%%%%%%%%%%%%%%%%%%%%%%%%%%%%%%%%%%%%%%%%%%%%%%
%%%%%%%%%%%%%%%%%%%%%%%%%%%%%%%%%%%%%%%%%%%%%%%%%%%%%%%%%%%%%%%%%%%%%%

In this section we are interested in the case in which the target-space metric
$g_{xy}(\phi)$ of the non-linear $\sigma$-model action
Eq.~(\ref{eq:sigmamodelaction}) is that of a $G/H$ coset space which is a
Riemannian symmetric space.\footnote{In this section we are going to use, with
  minimal changes, the notation and conventions of
  Refs.~\cite{Ortin:2015hya,Bandos:2016smv} to which we refer for further
  references and details on symmetric $\sigma$-models.} In particular, we are
going to assume that $g_{xy}(\phi)$ has been constructed using the restriction
of the Killing metric of $G$, $g_{AB}$, to the horizontal space,\footnote{This
  vector space, $\mathfrak{t}$ is the complement of the Lie subalgebra
  $\mathfrak{h}$ of H in $\mathfrak{g}$ (the Lie algebra of $G$), that is,
  $\mathfrak{g}=\mathfrak{h}\oplus \mathfrak{t}$. We use indices
  $A,B,\dots=1,\ldots,$dim\,$G$ to label the adjoint representation of $G$,
  indices $i,j,\ldots=1,\ldots,$dim\,$H$ to label that of $H$ and
  $m,n,\ldots=1,\ldots,$dim\,$G$-dim\,$H$ to label a basis of
  $\mathfrak{t}$. The scalars are labeled by
  $x,y,\ldots =1,\ldots,$dim\,$G$-dim\,$H$.}  $g_{mn}$ and the horizontal
components of the Maurer-Cartan 1-form $v^{m}=v^{m}{}_{x}d\phi^{x}$ as
Vielbeins:

\begin{equation}
g_{xy} = g_{mn}v^{m}{}_{x}v^{n}{}_{y}\,. 
\end{equation}

\noindent
Thus, $g_{xy}$ admits, at least,\footnote{We will ignore, for the sake of
  simplicity, any other Killing vectors of $g_{xy}$.} dim\,$G$ Killing
vectors, $k_{A}{}^{x}$, which generate as many global symmetries of the action
Eq.~(\ref{eq:transformationscalars}). Associated to them, there are dim\,$G$
closed $(d-1)$-form currents, $J_{A}$, of the form
Eq.~(\ref{eq:NoethercurrentsjA}) and, by construction, there are more
conserved currents (dim\,$G$) than physical scalars
(dim\,$G$-dim\,$H$). However, as noticed in Ref.~\cite{Bandos:2016smv}, only
if we use all of them will the whole global symmetry group, $G$, be
preserved. Therefore, we must define dim\,$G$ $(d-2)$-forms $C_{A}$ and their
respective field strengths $G_{A}$, through Eq.~(\ref{eq:GAstarjA}) and we
must use all of them in the action, but we must find a way to make dim\,$H$ of
the $(d-2)$-forms $C_{A}$ non-dynamical.

On the other hand, as we have discussed, it is clear that it is impossible to
construct an equivalent action in which only the dual $(d-2)$-forms $C_{A}$,
and not the scalar fields $\phi^{x}$ occur. The best we can hope for is a
democratic pseudoaction.

Taking into account all this and the discussions in
Ref.~\cite{Bandos:2016smv}, we propose the following democratic action for all
these fields:

\begin{equation}
  \begin{aligned}
    S_{\rm Dem}[e^{a},\phi^{x},C_{A}]
    & =
    \int \left\{
    (-1)^{d-1} \star (e^{a}\wedge e^{b}) \wedge R_{ab}
      +\tfrac{1}{4}g_{xy}d\phi^{x}\wedge d\phi^{y}
      \right.
      \\
      & \\
      & \hspace{.5cm}
      \left.
      +\tfrac{(-1)^{d}}{4}\mathfrak{M}^{AB}G_{A}\wedge \star G_{B}
      -\tfrac{(-1)^{d}}{2}g^{AB}G_{A}\wedge \hat{k}_{B} \right\}\,,
  \end{aligned}
\end{equation}

\noindent
where the dim\,$G\times$dim\,$G$ matrix $\mathfrak{M}^{AB}$ is defined as

\begin{equation}
  \label{eq:mathfrakMABdef}
  \mathfrak{M}^{AB}
  =
  g^{AC}g^{BD}k_{C}{}^{x}k_{D}{}^{y}g_{xy}
  % =
  % g^{AC}g^{BD}k_{C}{}^{m}k_{D}{}^{n}g_{mn}
  \,,
\end{equation}

\noindent
and where the equations of motion are meant to be supplemented by the duality
relations Eqs.~(\ref{eq:GAstarjA}).

It is not difficult to prove that the matrix $\mathfrak{M}^{AB}$ has
dim\,$G$-dim\,$H$ eigenvectors so that
rank\,$\mathfrak{M} =\,$dim\,$G$-dim\,$H$.\footnote{The eigenvectors are,
  precisely, the dim\,$H$ \textit{momentum maps}
  $P_{A}{}^{i}= \Gamma_{\rm Adj}(u^{-1})^{i}{}_{A}$ \cite{Bandos:2016smv}:
  taking into account that the Killing vectors are given by
  $k_{A}{}^{m}= -\Gamma_{\rm Adj}(u^{-1})^{m}{}_{A}$ and that the Killing
  metric is $g$-invariant and block-diagonal,
  \begin{equation}
    \mathfrak{M}^{AB} P_{B}{}^{i}
    =
    -g^{AC}k_{C}{}^{m}g_{mn} \Gamma_{\rm Adj}(u^{-1})^{n}{}_{A}g^{BD}
    \Gamma_{\rm Adj}(u^{-1})^{i}{}_{A}
    =
    -g^{AC}k_{C}{}^{m}g_{mn} g^{ni}
    =
    0\,.
  \end{equation}
  On the other hand, using the same properties we can show that
  \begin{equation}
    \label{eq:Mk=gk}
    \mathfrak{M}^{AB} k_{B}{}^{m}
    =
    g^{AB}k_{B}{}^{m}\,.
  \end{equation}
  These results are compatible because in $G/H$ only dim\,$G-$ dim\,$H$
  vectors are linearly independent at any given point.
}This means that in the above action there are dim\,$H$ combinations of the
magnetic field strengths $G_{A}$ which do not occur in the kinetic term, which
is what we need in order not to have too many dynamical fields.

We are going to show that the equations of motion that follow from the above
democratic action are those of the original $\sigma$-model upon use of the
duality relations Eqs.~(\ref{eq:GAstarjA}).

Observe that the original kinetic term for the scalar fields can be rewritten
in terms of the Noether 1-form currents using the matrix $\mathfrak{M}$ as

\begin{equation}
  \mathfrak{M}^{AB}\star J_{A}\wedge J_{B}
  =
  g^{AC}g^{BD}k_{C}{}^{x}k_{D}{}^{y}g_{xy} k_{A\,z}k_{B\,w}
  d\phi^{z}\wedge\star d\phi^{w}
  =
  g_{xy}d\phi^{x}\wedge \star d\phi^{y}\,,  
\end{equation}

\noindent
by virtue of the property

\begin{equation}
    \label{eq:prodK}
  g^{AB}k_{A}{}^{m}k_{B}{}^{n} = g^{mn}\,,
  \,\,\,\,\,\,
  \Rightarrow
  \,\,\,\,\,\,
  g^{AB}k_{A}{}^{x}k_{B}{}^{y} = g^{xy}\,.
\end{equation}

Observe that, then, 

\begin{equation}
  \label{eq:Mkk=gkk=g}
  \mathfrak{M}^{AB}k_{A\, x}k_{B\, y}
  =
  g^{AB}k_{A\, x}k_{B\, y}
  =
  g_{xy}\,.
\end{equation}

The Einstein equations that follow from the democratic action are

\begin{equation}
  \begin{aligned}
    \label{eq:Eadem}
  \mathbf{E}_{a}
  & =
    \imath_{a}\star (e^{c}\wedge e^{d})\wedge R_{cd}
      +\tfrac{(-1)^{d}}{4}g_{xy}
      \left(\imath_{a}d\phi^{x}\star d\phi^{y}
      +d\phi^{x}\wedge \imath_{a}\star d\phi^{y}\right)
    \\
    & \\
    &\hspace{.5cm}
    +\tfrac{1}{4}\mathfrak{M}^{AB}
      \left(\imath_{a}G_{A}\wedge \star G_{B}
      +(-1)^{d}G_{A}\imath_{a}\star G_{B}\right)\,.
  \end{aligned}
\end{equation}

It is enough to consider the last term (the energy-momentum $(d-1)$-form of
the dual $(d-2)$-forms). Using the duality relations Eqs.~(\ref{eq:GAstarjA})
and the property Eq.~(\ref{eq:Mk=gk}), that term takes the form

\begin{equation}
  \begin{aligned}
    \tfrac{1}{4}\mathfrak{M}^{AB} \left(\imath_{a}\star \hat{k}_{A}\wedge
      \hat{k}_{B} +(-1)^{d}\star \hat{k}_{A} \imath_{a}\hat{k}_{B}\right)
    & =
    \tfrac{(-1)^{d}}{4}\mathfrak{M}^{AB} \left( \hat{k}_{A}\wedge
      \imath_{a}\star \hat{k}_{B} +\imath_{a}\hat{k}_{A}\star \hat{k}_{B}
    \right)
    \\
    & \\
    & =
    \tfrac{(-1)^{d}}{4}\mathfrak{M}^{AB}k_{A\, x}k_{B\, y}
    \left( d\phi^{x}\wedge \imath_{a}\star d\phi^{y}
      +\imath_{a} d\phi^{x}\star  d\phi^{y}
    \right)
    \\
    & \\
    & =
    \tfrac{(-1)^{d}}{4}g_{xy}
    \left( d\phi^{x}\wedge \imath_{a}\star d\phi^{y}
      +\imath_{a} d\phi^{x}\star  d\phi^{y}
    \right)\,,
  \end{aligned}
\end{equation}

\noindent
by virtue of Eq.~(\ref{eq:Mkk=gkk=g}) and it can be added to the
energy-momentum $(d-1)$-form of the scalars to recover the energy-momentum
tensor of the scalars in the original theory.

The equations of motion of the scalars are, after use of the Killing equation

\begin{equation}
  \label{eq:eomscalardemocraticaction}
  \begin{aligned}
    \mathbf{E}_{x}
    & =
    -\tfrac{1}{2}g_{xy}\left\{d\star d\phi^{y}
      +\Gamma_{zw}{}^{y}d\phi^{z}\wedge\star d\phi^{w}\right\}
    +\tfrac{(-1)^{d}}{4}\partial_{x}\mathfrak{M}^{AB}G_{A}\wedge \star G_{B}
    \\
    & \\
    & \hspace{.5cm}
    -\tfrac{1}{2}g^{AB}k_{A\, x}dG_{B}
    +(-1)^{d+1}g^{AB}\nabla_{x}k_{A\, y} G_{B}\wedge d\phi^{y}\,.
  \end{aligned}
\end{equation}

Let us consider the last term first.  Using the duality relations
Eqs.~(\ref{eq:GAstarjA})

\begin{equation}
  (-1)^{d+1}g^{AB}\nabla_{x}k_{A\, y} k_{B\, z}\star d\phi^{z}\wedge d\phi^{y}
  =
  \tfrac{(-1)^{d+1}}{2}\nabla_{x}g_{yz}\star d\phi^{z}\wedge d\phi^{y}
  =
0\,,
\end{equation}

\noindent
since we are using the target-space Levi-Civita connection.  The second term
can be put in the form

\begin{equation}
  \tfrac{(-1)^{d}}{2}g^{AC}g^{BD}\nabla_{x}k_{C\,y}k_{D}{}^{y}
  k_{A\, z}k_{B\, w}\star d\phi^{z} \wedge d\phi^{w}
  =
  \tfrac{(-1)^{d}}{2}g^{AC}\nabla_{x}k_{C\,w} 
  k_{A\, z}\star d\phi^{z} \wedge d\phi^{w}
  =
  0\,,
\end{equation}

\noindent
for the same reason.

The third (next to last) term in Eq.~(\ref{eq:eomscalardemocraticaction})
vanishes on account of the Bianchi identity of $G_{A}$, which is related to
the equation of motion of the scalars. Thus, instead of throwing it away, we
are going to use the duality relation and 
Eq.~(\ref{eq:eomdja}) in it

\begin{equation}
  \begin{aligned}
    -\tfrac{1}{2}g^{AB}k_{A\, x}dG_{B}
    & =
    -\tfrac{1}{2}g^{AB}k_{A\, x}dJ_{B}
    \\
    & \\
    & =
    -\tfrac{1}{2}g^{AB}k_{A\, x}k_{B\, y}\left\{d\star d\phi^{y}
      +\Gamma_{zw}{}^{y}d\phi^{z}\wedge\star d\phi^{w}\right\}
    \\
    & \\
    & =
    -\tfrac{1}{2}g_{xy}\left\{d\star d\phi^{y}
      +\Gamma_{zw}{}^{y}d\phi^{z}\wedge\star d\phi^{w}\right\}\,,
  \end{aligned}
\end{equation}

\noindent
and we recover the original equation of motion of the scalars with identical
normalization.

Finally, the equations of motion of the $(d-2)$-form potentials $C_{A}$

\begin{equation}
  \mathbf{E}^{A}
  =
  \tfrac{1}{2}d\left[\mathfrak{M}^{AB}\star G_{B}-g^{AB}\hat{k}_{B}\right]\,,
\end{equation}

\noindent
are solved automatically by the duality relation upon use of the properties of
the matrix $\mathfrak{M}^{AB}$.

The scalars of all the maximal and half-maximal supergravities parametrize a
Riemannian symmetric $\sigma$-model. However, in all those theories they are
also coupled to other fields. In the next section we consider the coupling to
$(p+1)$-form potentials as a toy model since ``real'' supergravities usually
have several of these with different ranks and with Chern-Simons terms in the
action and field strengths.

%%%%%%%%%%%%%%%%%%%%%%%%%%%%%%%%%%%%%%%%%%%%%%%%%%%%%%%%%%%%%%%%%%%%%% 
%%%%%%%%%%%%%%%%%%%%%%%%%%%%%%%%%%%%%%%%%%%%%%%%%%%%%%%%%%%%%%%%%%%%%%
%%%%%%%%%%%%%%%%%%%%%%%%%%%%%%%%%%%%%%%%%%%%%%%%%%%%%%%%%%%%%%%%%%%%%%
%%%%%%%%%%%%%%%%%%%%%%%%%%%%%%%%%%%%%%%%%%%%%%%%%%%%%%%%%%%%%%%%%%%%%%
\section{Dualization of Riemannian symmetric $\sigma$-models coupled to
  $(p+1)$-forms}
\label{sec-sigmamodelsymmetriccoupledtop+1forms}
%%%%%%%%%%%%%%%%%%%%%%%%%%%%%%%%%%%%%%%%%%%%%%%%%%%%%%%%%%%%%%%%%%%%%% 
%%%%%%%%%%%%%%%%%%%%%%%%%%%%%%%%%%%%%%%%%%%%%%%%%%%%%%%%%%%%%%%%%%%%%%
%%%%%%%%%%%%%%%%%%%%%%%%%%%%%%%%%%%%%%%%%%%%%%%%%%%%%%%%%%%%%%%%%%%%%%
%%%%%%%%%%%%%%%%%%%%%%%%%%%%%%%%%%%%%%%%%%%%%%%%%%%%%%%%%%%%%%%%%%%%%%

The next step consists in the coupling of a Riemannian symmetric
$\sigma$-model to a set of $(p+1)$-form potentials (the fields $p$-branes
naturally couple to)

\begin{equation}
  A^{\Lambda}
  =
  \frac{1}{(p+1)!}A^{\Lambda}{}_{\mu_{1}\cdots\mu_{p+1}}dx^{\mu_{1}}\wedge
  \cdots \wedge dx^{\mu_{p+1}}\,,
\end{equation}

\noindent
with $(p+2)$-form field strengths

\begin{equation}
F^{\Lambda} = dA^{\Lambda}\,,  
\end{equation}

\noindent
invariant under the gauge transformations

\begin{equation}
  \label{eq:deltachiALambda}
  \delta_{\chi}A^{\Lambda}
  =
  d\chi^{\Lambda}\,,
\end{equation}

\noindent
where each $\chi^{\Lambda}$ is an arbitrary $p$-form.

In arbitrary dimension $d$, the action that describes this coupling takes the
generic form

\begin{equation}
  \label{eq:sigmamodelcoupledtop+1formsaction}
  S[e^{a},A^{\Lambda},\phi^{x}]
  =
  \int
  \left\{
    (-1)^{d-1} \star (e^{a}\wedge e^{b}) \wedge R_{ab}
    +\tfrac{1}{2}g_{xy}d\phi^{x}\wedge \star d\phi^{y}
    -\tfrac{(-1)^{(p+1)d}}{2}I_{\Lambda\Sigma}F^{\Lambda}\wedge \star F^{\Sigma}
  \right\}\,,  
\end{equation}

\noindent
$I_{\Lambda\Sigma}$ being a symmetric and negative-definite scalar-dependent
matrix.

The equations of motion that follow from this action are

\begin{subequations}
  \begin{align}
    \mathbf{E}_{a}
    & =
          \imath_{a}\star (e^{c}\wedge e^{d})\wedge R_{cd}
      +\tfrac{(-1)^{d}}{2}g_{xy}
    \left(\imath_{a}d\phi^{x}\wedge \star d\phi^{y}
      +d\phi^{x}\wedge \imath_{a}\star d\phi^{y}\right)
      \nonumber \\
    & \nonumber \\
    & \hspace{.5cm}
      +\tfrac{(-1)^{(p+1)d}}{2} I_{\Lambda\Sigma}
      \left(
      \imath_{a}F^{\Lambda}\wedge \star F^{\Sigma}
      +(-1)^{p+1}F^{\Lambda}\wedge \imath_{a}\star F^{\Sigma}
    \right)\,,
    \\
    & \nonumber \\
        \mathbf{E}_{x}
    & =
      -g_{xy}\left[d\star d\phi^{y}
      +\Gamma_{zw}{}^{y}d\phi^{z}\wedge \star d\phi^{w}\right]
      -\tfrac{(-1)^{(p+1)d}}{2}\partial_{x}I_{\Lambda\Sigma}
      F^{\Lambda}\wedge \star F^{\Sigma}\,,
    \\
    & \nonumber \\
        \mathbf{E}_{\Lambda}
    & =
      d\left(I_{\Lambda\Omega}\star F^{\Omega}\right)\,.
    % \\
    % & \nonumber \\
    % \mathbf{\Theta}
    % & =
    %   - \star (e^{a}\wedge e^{b}) \wedge \delta \omega_{ab}
    %   +g_{xy}\star d\phi^{y} \delta\phi^{x}
    %   -\tfrac{(-1)^{(p+1)d}}{2} \delta A^{\Lambda}\wedge
    %   \left(I_{\Lambda\Sigma}\star F^{\Sigma}\right)\,.
  \end{align}
\end{subequations}

For a non-constant kinetic matrix $I_{\Lambda\Sigma}$ the action is invariant
under all the transformations of the scalars
Eq.~(\ref{eq:transformationscalars}) associated to the symmetries of the
$\sigma$-model, if the $(p+1)$-forms also transform according to

\begin{equation}
  \delta_{A}A^{\Lambda}
  =
  T_{A}{}^{\Lambda}{}_{\Sigma}A^{\Sigma}\,,
\end{equation}

\noindent
for some matrices $T_{A}$, and the kinetic matrix $I_{\Lambda\Sigma}$
satisfies the property

\begin{equation}
  \delta_{A}I_{\Lambda\Sigma}
  =
  -\pounds_{k_{A}}I_{\Lambda\Sigma}
  =
  -2T_{A}{}^{\Omega}{}_{(\Lambda}I_{\Sigma)\Omega}\,.
\end{equation}

The Noether current $(d-1)$-forms associated to these symmetries are

\begin{equation}
  J_{A}
  =
  \star \hat{k}_{A} +(-1)^{(p+1)d}T_{A}{}^{\Lambda}{}_{\Sigma}A^{\Sigma}\wedge
  \left(I_{\Lambda\Omega}\star F^{\Omega}\right)\,.
\end{equation}

The first term of this current is invariant under the gauge transformations 
Eq.~(\ref{eq:deltachiALambda}) but the second is not: it transforms into a
total derivative on-shell. Thus, if we try to dualize $J_{A}$ using its
conservation law $dJ_{A}=0$

\begin{equation}
\star \hat{k}_{A} +(-1)^{(p+1)d}T_{A}{}^{\Lambda}{}_{\Sigma}A^{\Sigma}\wedge
\left(I_{\Lambda\Omega}\star F^{\Omega}\right)
\equiv
d\tilde{C}_{A}\,,
\end{equation}

\noindent
the right definition for a gauge-invariant field strength is

\begin{equation}
\star \hat{k}_{A} 
=
d\tilde{C}_{A}-(-1)^{(p+1)d}T_{A}{}^{\Lambda}{}_{\Sigma}A^{\Sigma}\wedge
\left(I_{\Lambda\Omega}\star F^{\Omega}\right)
\equiv
G_{A}\,,
\end{equation}

\noindent
and the total derivative generated by the gauge transformations of the
$(p+1)$-form potentials $A^{\Lambda}$ must be absorbed by a gauge
transformation of the $(d-2)$-form potentials that we will described shortly.

The Chern-Simons term in the field strength is unusual but can be transformed
using the dual of the $(p+1)$-form potentials: their equations of motion
$\mathbf{E}_{\Lambda}=0$ can be locally solved with the introduction of the
dual $(\tilde{p}+1)$-forms $\tilde{A}_{\Lambda}$, with
$\tilde{p}\equiv d-p-4$:

\begin{equation}
  I_{\Lambda\Omega}\star F^{\Omega}
  \equiv
  d\tilde{A}_{\Lambda}
  \equiv
  \tilde{F}_{\Lambda}\,.
\end{equation}

The field strengths $\tilde{F}_{\Lambda}$ are invariant under the dual gauge
transformations

\begin{equation}
  \delta_{\tilde{\chi}} \tilde{A}_{\Lambda}
  =
  d\tilde{\chi}_{\Lambda}\,.
\end{equation}

Using this definition, we can write

\begin{equation}
  G_{A}
=
d\tilde{C}_{A}
-(-1)^{(p+1)d}T_{A}{}^{\Lambda}{}_{\Sigma}A^{\Sigma}\wedge \tilde{F}_{\Lambda}\,,
\end{equation}

\noindent
and, integrating by parts in order to get a more symmetric expression, we
arrive at the final definition of the $(d-2)$-form potentials and their field
strengths dual to the scalars:

\begin{equation}
G_{A}
\equiv
dC_{A}
-\tfrac{(-1)^{(p+1)d}}{2}T_{A}{}^{\Lambda}{}_{\Sigma}
\left(A^{\Sigma}\wedge \tilde{F}_{\Lambda}
  +(-1)^{p(d+1)}\tilde{A}_{\Lambda}\wedge F^{\Sigma}\right)
=
\star \hat{k}_{A}\,.
\end{equation}

The gauge invariance of the field strengths $G_{A}$ implies the following
gauge transformations of the $(d-2)$-form potentials

\begin{equation}
  \delta C_{A}
  =
  d\Sigma_{A}
  +\tfrac{(-1)^{(p+1)d}}{2}T_{A}{}^{\Lambda}{}_{\Sigma}
\left(\chi^{\Sigma}\wedge \tilde{F}_{\Lambda}
  +(-1)^{p(d+1)}\tilde{\chi}_{\Lambda}\wedge F^{\Sigma}\right)\,.
\end{equation}

Observe that in this theory

\begin{equation}
  k_{A}{}^{x}\mathbf{E}_{x}
  =
  -dJ_{A}
  +(-1)^{(p+1)(d+1)} T_{A}{}^{\Lambda}{}_{\Sigma}A^{\Sigma}\wedge \mathbf{E}_{\Lambda}\,.
\end{equation}

Based on our previous experience, we propose the following gauge- and
$G$-invariant democratic action for this theory:

\begin{equation}
  \begin{aligned}
    S_{\rm Dem}[e^{a},A^{\Lambda},\tilde{A}_{\Lambda},\phi^{x},C_{A}]
    & =
    \int \left\{
    (-1)^{d-1} \star (e^{a}\wedge e^{b}) \wedge R_{ab}
      +\tfrac{1}{4}g_{xy}d\phi^{x}\wedge d\phi^{y}
      \right.
      \\
      & \\
      & \hspace{.5cm}
       -\tfrac{(-1)^{(p+1)d}}{4}I_{\Lambda\Sigma}F^{\Lambda}\wedge \star F^{\Sigma}
       -\tfrac{(-1)^{(\tilde{p}+1)d}}{4}I^{\Lambda\Sigma}
       \tilde{F}_{\Lambda}\wedge \star \tilde{F}_{\Sigma}
      \\
      & \\
      & \hspace{.5cm}
      \left.
      +\tfrac{(-1)^{d}}{4}\mathfrak{M}^{AB}G_{A}\wedge \star G_{B}
      -\tfrac{(-1)^{d}}{2}g^{AB}G_{A}\wedge \hat{k}_{B} \right\}\,,
  \end{aligned}
\end{equation}

\noindent
where

\begin{equation}
  I^{\Lambda\Omega}I_{\Omega\Sigma}
  =
  \delta^{\Lambda}{}_{\Sigma}\,,
\end{equation}

\noindent
where we must use the duality relations

\begin{subequations}
  \begin{align}
    I_{\Lambda\Omega}\star F^{\Omega}
    & =
    \tilde{F}_{\Lambda}\,,
    \\
    & \nonumber \\
    \star \hat{k}_{A}
    & =
      G_{A}\,,
  \end{align}
\end{subequations}

\noindent
in the equations of motion in order to recover those of the original
theory,\footnote{Those of the $(p+1)$-form potentials appear with a factor of
  $1/2$.}  as it can easily be checked.

Coupling the scalars to more potentials of different ranks should only involve
more Chern-Simons terms in the definition of the dual field strengths
$G_{A}$. However, there can be additional complications, as we will see in the
case of the $\mathcal{N}=2B,d=10$ theory.

On the other hand, when $d=2(p+2)$ new couplings between the scalars and
$(p+1)$-form potentials are possible. This is a specially interesting case
because it includes all the $\mathcal{N}\geq 3,d=4$ ungauged supergravities
and because some of the symmetries of the theory (electric-magnetic dualities)
are realized as symmetries of the equations of motion only. We consider it
next.

%%%%%%%%%%%%%%%%%%%%%%%%%%%%%%%%%%%%%%%%%%%%%%%%%%%%%%%%%%%%%%%%%%%%%%
%%%%%%%%%%%%%%%%%%%%%%%%%%%%%%%%%%%%%%%%%%%%%%%%%%%%%%%%%%%%%%%%%%%%%%
%%%%%%%%%%%%%%%%%%%%%%%%%%%%%%%%%%%%%%%%%%%%%%%%%%%%%%%%%%%%%%%%%%%%%%
%%%%%%%%%%%%%%%%%%%%%%%%%%%%%%%%%%%%%%%%%%%%%%%%%%%%%%%%%%%%%%%%%%%%%%
\subsection{The $d=2(p+2)$ case and electric-magnetic dualities}
\label{sec-electricmagnetic}
%%%%%%%%%%%%%%%%%%%%%%%%%%%%%%%%%%%%%%%%%%%%%%%%%%%%%%%%%%%%%%%%%%%%%% 
%%%%%%%%%%%%%%%%%%%%%%%%%%%%%%%%%%%%%%%%%%%%%%%%%%%%%%%%%%%%%%%%%%%%%%
%%%%%%%%%%%%%%%%%%%%%%%%%%%%%%%%%%%%%%%%%%%%%%%%%%%%%%%%%%%%%%%%%%%%%%
%%%%%%%%%%%%%%%%%%%%%%%%%%%%%%%%%%%%%%%%%%%%%%%%%%%%%%%%%%%%%%%%%%%%%%

%%%%%%%%%%%%%%%%%%%%%%%%%%%%%%%%%%%%%%%%%%%%%%%%%%%%%%%%%%%%%%%%%%%%%%
%%%%%%%%%%%%%%%%%%%%%%%%%%%%%%%%%%%%%%%%%%%%%%%%%%%%%%%%%%%%%%%%%%%%%%
%%%%%%%%%%%%%%%%%%%%%%%%%%%%%%%%%%%%%%%%%%%%%%%%%%%%%%%%%%%%%%%%%%%%%%
%%%%%%%%%%%%%%%%%%%%%%%%%%%%%%%%%%%%%%%%%%%%%%%%%%%%%%%%%%%%%%%%%%%%%%
\subsubsection{The theory and its dualities}
%%%%%%%%%%%%%%%%%%%%%%%%%%%%%%%%%%%%%%%%%%%%%%%%%%%%%%%%%%%%%%%%%%%%%% 
%%%%%%%%%%%%%%%%%%%%%%%%%%%%%%%%%%%%%%%%%%%%%%%%%%%%%%%%%%%%%%%%%%%%%%
%%%%%%%%%%%%%%%%%%%%%%%%%%%%%%%%%%%%%%%%%%%%%%%%%%%%%%%%%%%%%%%%%%%%%%
%%%%%%%%%%%%%%%%%%%%%%%%%%%%%%%%%%%%%%%%%%%%%%%%%%%%%%%%%%%%%%%%%%%%%%

When $d = 2(p+2)$ it is possible to add a gauge-invariant topological
(metric-independent) term to the action
Eq.~(\ref{eq:sigmamodelcoupledtop+1formsaction}), which takes the generic form

\begin{equation}
  \label{eqAction}
  S[e^{a},A^{\Lambda},\phi^{x}]
  =
  \int\left\{ - \star (e^{a} \wedge e^{b} ) \wedge R_{ab}
    +\tfrac{1}{2}g_{xy}d\phi^{x} \wedge \star d\phi^{y}
    - \tfrac{1}{2} I_{\Sigma \Lambda} F^{\Sigma}
    \wedge \star F^\Lambda
    -\tfrac{1}{2}R_{\Sigma\Lambda} F^{\Sigma} \wedge F^{\Lambda}
    \right\}\,,
\end{equation}

\noindent
where the new matrix $R_{\Sigma \Lambda}$ also depends on the scalar fields.

While $I_{\Lambda\Sigma}$ is always symmetric (and, conventionally,
negative-definite) the symmetry properties of $R_{\Sigma \Lambda}$ depend on
the dimension $d$:\footnote{One should also take into account that, for
  $(p+2)$-forms in $d=2(p+2)$ dimensions
  \begin{subequations}
    \begin{align}
      \star^{2}F^{\Lambda}
      & =
        \xi^{2}F^{\Lambda}\,,
      \\
      & \nonumber \\
      F^{\Lambda}\wedge F^{\Sigma}
      & =
      -\xi^{2}F^{\Sigma}\wedge F^{\Lambda}\,,
      \\
      & \nonumber \\
      \star F^{\Lambda}\wedge \star F^{\Sigma}
      & =
       -F^{\Lambda}\wedge  F^{\Sigma}\,.
    \end{align}
  \end{subequations}
}

\begin{equation}
  R^{T} = - \xi^{2} R \,,
  \hspace{1cm}
  \xi^{2} = (-1)^{p+1} \,.
\end{equation}

The equations of motion that follow from this action are

\begin{subequations}
  \label{eqmoto1}
\begin{align}
  \mathbf{E}_{c}
  & =
    \imath_{c} \star \left(e^{a} \wedge e^{b}\right) \wedge R_{ab}
    +\tfrac{1}{2} g_{xy} \left(\imath_{c} d\phi^{x} \wedge \star d\phi^{y} +
    d\phi^{x} \wedge \imath_{c} \star d\phi^{y} \right)
    \nonumber \\
  & \nonumber \\
  &\hspace{.5cm}
    -\tfrac{1}{2} I_{\Sigma\Lambda} \left(\imath_{c} F^{\Sigma} \wedge \star
    F^{\Lambda}
    +\xi^{2} F^{\Sigma} \wedge \imath_{c} \star F^{\Lambda}\right) \,,
  \\
  & \nonumber \\
  \label{eq:Ez}
  \mathbf{E}_{z}
  & =
    -g_{zw} \left[d\star d\phi^{w}
      +\Gamma_{xy}{}^{w}d\phi^{x}\wedge \star d\phi^{y}\right]
    -\tfrac{1}{2}\partial_{z} I_{\Sigma\Lambda} F^{\Sigma}\wedge\star F^{\Lambda}
    -\tfrac{1}{2}\partial_{z} R_{\Sigma\Lambda} F^{\Sigma} \wedge F^{\Lambda} \,,
  \\
  & \nonumber \\
  \mathbf{E}_{\Sigma}
  & =
    d \left(I_{\Sigma \Lambda} \star F^{\Lambda}
    +R_{\Sigma \Lambda} F^{\Lambda}\right)\,,
%   \\
%   & \nonumber \\
%   \mathbf{\Theta}
%   & = -\star \left(e^{a} \wedge e^{b} \right) \wedge \delta \omega_{ab}
%     -I_{\Sigma\Lambda} \delta A^{\Sigma} \wedge \star F^{\Lambda}
%     -R_{\Sigma\Lambda} \delta A^{\Sigma} \wedge F^{\Lambda}
%     \nonumber \\
%   & \nonumber \\
% & \hspace{.5cm} + g_{xy} \delta \phi^{x} \star d\phi^{y}  \,,
\end{align}
\end{subequations}

The equations of motion of the $(p+1)$-forms $A^{\Lambda}$ can be locally
solved by introducing dual $(p+1)$-forms $A_{\Lambda}$ such that

\begin{equation}
I_{\Sigma \Lambda} \star F^{\Lambda}
+R_{\Sigma \Lambda} F^{\Lambda}
=
dA_{\Lambda}
\equiv
F_{\Lambda}\,,
\end{equation}
 
\noindent
where $F_{\Lambda}$ are the associated $(p+2)$-form field strengths. it is,
then, natural, to introduce

\begin{equation}
  \left(A^{M}\right)
  \equiv
  \left(
    \begin{array}{c}
      A^{\Lambda} \\ A_{\Lambda} \\
    \end{array}
  \right)\,,
  \hspace{1cm}
  \left(F^{M}\right)
  \equiv
  d\left(A^{M}\right)
  \equiv
  \left(
    \begin{array}{c}
      F^{\Lambda} \\ F_{\Lambda} \\
    \end{array}
  \right)\,.
\end{equation}

The dual field strengths $F_{\Lambda}$ have been defined in terms of the
original ones $F^{\Lambda}$ and the scalars. Therefore, it is not surprising
that $F^{M}$ satisfies the so-called \textit{twisted self-duality constraint}

\begin{equation}
  \label{eq:twistedselfduality}
  \star F^{M}
  =
  \xi^{2} \Omega^{MN} \mathcal{M}_{NP} F^{P}\,,
\end{equation}

\noindent
where we have defined

\begin{equation}
  \Omega
  \equiv
  \left(\Omega_{MN}\right)
  \equiv
  \left(
    \begin{array}{lr}
      0 & \delta_{\Lambda}{}^{\Sigma} \\
      \xi^{2}\delta^{\Lambda}{}_{\Sigma} & 0 \\
    \end{array}
  \right)\,,
  \hspace{1cm}
  \Omega^{-1}
  \equiv
  \left(\Omega^{MN}\right)
  \equiv
  \left(
    \begin{array}{lr}
      0 & \xi^{2}\delta^{\Lambda}{}_{\Sigma} \\
      \delta_{\Lambda}{}^{\Sigma} & 0 \\
    \end{array}
  \right)
  =
  \xi^{2}\Omega\,,
\end{equation}

\noindent
which, for $\xi^{2}=-1$ is the Sp$(2n,\mathbb{R})$ metric and for $\xi^{2}=+1$
is the off-diagonal O$(n,n)$ metric, and the symmetric scalar matrix

\begin{equation}
  \mathcal{M}
  =
  \left( \mathcal{M}_{MN}\right)
  = \begin{pmatrix}
      I_{\Lambda\Sigma}-\xi^{2}R_{\Lambda\Gamma}I^{\Gamma\Omega}R_{\Omega\Sigma}\,\,
      &
        \xi^{2} R_{\Lambda\Gamma}I^{\Gamma\Sigma}
      \\
      & \\
      -I^{\Lambda \Gamma}R_{\Gamma\Sigma}
      &
        I^{\Lambda\Sigma} 
    \end{pmatrix}
    =
    \begin{pmatrix}
      I - \xi^{2} R I^{-1} R\,\, & \xi^{2} R I^{-1} \\
                             & \\
      - I^{-1}R & I^{-1} \\
\end{pmatrix}\,,
\end{equation}

\noindent
which is symplectic for $\xi^{2}=-1$ or orthogonal for  $\xi^{2}=+1$
because

\begin{equation}
  \label{eq:MpreservesOmega}
  \mathcal{M}^{-1\, T}\Omega\mathcal{M}^{-1}
  =
  \Omega\,.
\end{equation}

The equations of motion of the $(p+1)$-forms $\mathbf{E}_{\Lambda}$ and the
Bianchi identities of their $(p+2)$-form field strengths
$\mathbf{B}^{\Lambda}=dF^{\Lambda}$ can be written in a compact way as
Bianchi identities for $F^{M}$

\begin{equation}
  \label{eq:MaxwellplusBianchis}
  d\left(F^{M}\right)
  =
  \left(
    \begin{array}{c}
      \mathbf{B}^{\Lambda} \\ \mathbf{E}_{\Lambda} \\
    \end{array}
  \right)
  =
  0\,.
\end{equation}

These equations are invariant under GL$(2n,\mathbb{R})$ transformations

\begin{equation}
  F^{M\,\prime}  = S^{M}{}_{N} F^{N}\,,
  \,\,\,\,\,\,
  \text{or}
  \,\,\,\,\,\,
  \mathcal{F}' =\mathcal{S}\mathcal{F}\,,
\end{equation}

\noindent
but the twisted self-duality constraint Eq.~(\ref{eq:twistedselfduality}) is
only invariant if, at the same time, the scalar matrix $\mathcal{M}$
transforms as

\begin{equation}
\mathcal{M}^{\prime}
  =
 \left(\Omega \mathcal{S} \Omega^{-1}\right)\mathcal{M}\mathcal{S}^{-1}\,.  
\end{equation}

\noindent
This implies that the scalar fields must transform as well.

Since the energy-momentum tensor of the $(p+1)$-form potentials can be written
in the form

\begin{equation}
  \label{eq:emtensorforms}
      -\tfrac{1}{2} I_{\Sigma\Lambda} \left(\imath_{c} F^{\Sigma} \wedge \star
    F^{\Lambda}
    +\xi^{2} F^{\Sigma} \wedge \imath_{c} \star F^{\Lambda}\right)
  =
  -\tfrac{1}{2}\Omega_{MN}\imath_{c}F^{M}\wedge F^{N}\,,
\end{equation}

\noindent
the Einstein equations will be invariant if

\begin{equation}
  \mathcal{S}^{T} \Omega \mathcal{S} = \Omega\,,
  \,\,\,\,\,
  \Rightarrow
  \,\,\,\,\,
  \Omega\mathcal{S}\Omega^{-1} = \mathcal{S}^{-1\, T}\,,
\end{equation}

\noindent
\textit{i.e.}~if $\mathcal{S}\in\,$Sp$(2n,\mathbb{R})$ when $\xi^{2}=-1$ ($p$
even, $d=4n$) and $\mathcal{S}\in\,$O$(n,n)$ when $\xi^{2}=+1$ ($p$ odd,
$d=4n+2$). This is a well-known generalization of the Gaillard-Zumino result
for $p=0$, presented in Ref.~\cite{Gaillard:1981rj}\footnote{See
  Refs.~\cite{deAntonioMartin:2012bi,Ortin:2015hya,Bandos:2016smv} and
  references therein.}.

Then, we conclude that, under symplectic or orthogonal rotations of the
potentials, $\mathcal{M}$ must transform as

\begin{equation}
  \label{eq:lineartransformationsM}
  \mathcal{M}^{\prime}
  =
  \mathcal{S}^{-1\, T}\mathcal{M}\mathcal{S}^{-1}\,,
\end{equation}

\noindent
for the twisted self-duality constraint to be respected. These rotations
preserve the energy-momentum tensor of the potentials. If we rewrite their
kinetic term in the action in the form

\begin{equation}
\sim \mathrm{Tr}\left(d\mathcal{M}^{-1}\wedge \star d\mathcal{M}\right)\,,
\end{equation}

\noindent
the invariance of this kinetic term and of the corresponding energy-momentum
tensor is manifest. However, this invariance is only apparent since we have
not yet described the action of these transformations on the scalar fields,
which only transform via field redefinitions or, equivalently, coordinate
transformations in the target space, which take the infinitesimal form

\begin{equation}
\delta_{\alpha} \phi^{x} = \alpha k^{x}(\phi)\,,
\end{equation}

\noindent
where $\alpha$ is an infinitesimal parameter and $k^{x}(\phi)$ is a target
space vector field. Since these transformations must preserve the kinetic
term, they must be Killing vectors of the target space metric $g_{xy}(\phi)$.
If $\{k_{A}{}^{x}\}$ is the set of these Killing vectors, the possible
transformations are

\begin{equation}
\delta_{\alpha} \phi^{x} = \alpha^{A} k_{A}{}^{x}(\phi)\,,
\end{equation}

\noindent
where now we have as many independent infinitesimal parameters $\alpha^{A}$ as
Killing vectors. These Killing vectors generate an isometry group $G$ which
is, in general smaller than Sp$(2n,\mathbb{R})$ or O$(n,n)$. 

These transformations act on the scalar matrix $\mathcal{M}$ as

\begin{equation}
  \delta_{\alpha}\mathcal{M}
  =
  \alpha^{A} k_{A}{}^{x}\partial_{x}\mathcal{M}\,,
\end{equation}

\noindent
and, according to the previous discussion, they will lead to an invariance of
the equations of motion if they are equivalent to the linear transformations
Eq.~(\ref{eq:lineartransformationsM}). Infinitesimally

\begin{subequations}
  \begin{align}
    \mathcal{S}
    & =
      1+\alpha^{A}T_{A}\,,
      \\
    & \nonumber \\
    \label{eq:infinitesimallineartransformationsM}
    \delta_{\alpha}\mathcal{M}
    & =
      -\alpha^{A}\left(T_{A}{}^{T}\mathcal{M}+\mathcal{M}T_{A}\right)\,,
  \end{align}
\end{subequations}

\noindent
where the matrices $T_{A}$ generate a representation of the isometry group $G$
through Sp$(2n,\mathbb{R})$ or O$(n,n)$ matrices.

The condition that the matrix $\mathcal{M}$ must satisfy for $G$ to be a
symmetry of the equations of motion and Bianchi identities of the $(p+1)$-form
potentials and of the Einstein equations is, therefore,

\begin{equation}
  \label{eq:equivarianceconditionM}
  k_{A}{}^{x}\partial_{x}\mathcal{M}
  +T_{A}{}^{T}\mathcal{M}+\mathcal{M}T_{A}
  =
  0\,.
\end{equation}

It only remains to show that these conditions are also sufficient for the
transformations to leave invariant the scalar equations of motion. First, we
need to rewrite them in a more symmetric form, using the matrix
$\mathcal{M}_{MN}$ and the vector of field strengths $F^{M}$. 

The simplest invariant that we can construct with these elements

\begin{equation}
-\tfrac{1}{4}\mathcal{M}_{MN}F^{M}\wedge \star F^{N}\,,
\end{equation}

\noindent
vanishes identically when we use the twisted self-duality constraint
Eq.~(\ref{eq:twistedselfduality}) and the preservation of $\Omega$ by
$\mathcal{M}$ Eq.~(\ref{eq:MpreservesOmega}):

\begin{equation}
  \begin{aligned}
  -\tfrac{1}{4}\mathcal{M}_{MN}F^{M}\wedge \star F^{N}
  & =
  -\tfrac{1}{4}\xi^{2}\mathcal{M}_{MN}\Omega^{NP}\mathcal{M}_{PQ}
    F^{M}\wedge F^{Q}
    \\
    & \\
  & =
  -\tfrac{1}{4}\Omega_{MQ} F^{M}\wedge F^{Q}\,,
  \end{aligned}
\end{equation}

\noindent
which vanishes identically because $\Omega_{MN}=\xi^{2}\Omega_{NM}$ while
$F^{M}\wedge F^{N} = -\xi^{2} F^{N}\wedge F^{M}$.

However,

\begin{equation}
  -\tfrac{1}{4}\partial_{z}\mathcal{M}_{MN}F^{M}\wedge \star F^{N}
  =
  -\tfrac{1}{4}\xi^{2}\partial_{z}\mathcal{M}_{MN}\Omega^{NP}\mathcal{M}_{PQ}
    F^{M}\wedge F^{Q}\,,
\end{equation}

\noindent
does not vanish because, by taking the derivative of
Eq.~(\ref{eq:MpreservesOmega}), one can easily see that

\begin{equation}
  \left(\partial_{z}\mathcal{M}\Omega^{-1}\mathcal{M} \right)^{T}
  =
  -\xi^{2}\partial_{z}\mathcal{M}\Omega^{-1}\mathcal{M}\,.
\end{equation}

A straightforward calculation shows that 

\begin{equation}
  -\tfrac{1}{4}\partial_{z}\mathcal{M}_{MN}F^{M}\wedge \star F^{N}
  =
  -\tfrac{1}{2}\partial_{z} I_{\Sigma\Lambda} F^{\Sigma}\wedge\star F^{\Lambda}
  -\tfrac{1}{2}\partial_{z} R_{\Sigma\Lambda} F^{\Sigma} \wedge F^{\Lambda}\,,
\end{equation}

\noindent
and we can rewrite the scalar equation of motion Eq.~(\ref{eq:Ez}) in the form

\begin{equation}
  \label{eq:Ez2}
  \mathbf{E}_{z}
   =
    -g_{zw} \left[d\star d\phi^{w}
      +\Gamma_{xy}{}^{w}d\phi^{x}\wedge \star d\phi^{y}\right]
      -\tfrac{1}{4}\partial_{z}\mathcal{M}_{MN}F^{M}\wedge \star F^{N}\,.
\end{equation}

Under the infinitesimal transformations

\begin{equation}
  \delta_{A}\phi^{x}
  =
  k_{A}{}^{x}\,,
  \hspace{1.5cm}
  \delta_{A}F
  =
  T_{A}F\,,
\end{equation}

\noindent
where $F\equiv (F^{M})$, the scalar equations of motion (\ref{eq:Ez2})
transform as

\begin{equation}
  \begin{aligned}
  \delta _{A}\mathbf{E}_{z}
    & =
      \partial_{z}k_{A}{}^{v}\mathbf{E}_{v}
    \\
    & \\
    & \hspace{.5cm}
      -2\nabla_{(z|}k_{A|w)}
      \left[d\star d\phi^{w}
     +\Gamma_{xy}{}^{w}d\phi^{x}\wedge \star d\phi^{y}\right]
    \\
    & \\
    & \hspace{.5cm}
    -g_{zw}
      \left(\nabla_{x}\nabla_{y}k^{w}+k^{v}R_{vxy}{}^{w}\right) 
      d\phi^{x}\wedge \star d\phi^{y}
    \\
    & \\
    & \hspace{.5cm}
      -\tfrac{1}{4}      \partial_{z}\left(
      k_{A}{}^{v}\partial_{v}\mathcal{M}_{PQ}+\mathcal{M}_{MQ}T_{A}{}^{M}{}_{P}
      +\mathcal{M}_{PN}T_{A}{}^{N}{}_{Q}\right)
      F^{P}\wedge \star F^{Q}\,.
  \end{aligned}
\end{equation}

The second and third lines vanish when $k_{A}{}^{x}$ is a Killing vector of
the target space metric, while the fourth vanishes upon use of the condition
Eq.~(\ref{eq:equivarianceconditionM}).

\subsubsection{Democratic pseudoaction I: the potentials}
%%%%%%%%%%%%%%%%%%%%%%%%%%%%%%%%%%%%%%%%%%%%%%%%%%%%%%%%%%%%%%%%%%%%%% 
%%%%%%%%%%%%%%%%%%%%%%%%%%%%%%%%%%%%%%%%%%%%%%%%%%%%%%%%%%%%%%%%%%%%%%
%%%%%%%%%%%%%%%%%%%%%%%%%%%%%%%%%%%%%%%%%%%%%%%%%%%%%%%%%%%%%%%%%%%%%%
%%%%%%%%%%%%%%%%%%%%%%%%%%%%%%%%%%%%%%%%%%%%%%%%%%%%%%%%%%%%%%%%%%%%%%

Using the results of the previous section it is not difficult to make an
educated guess for the democratic pseudoaction that contains the original and
dual potentials as independent variables: 

\begin{equation}
  \label{eq:democraticI}
  S[e^{a},A^{M},\phi^{x}]
  =
  \int\left\{-\star (e^{a}\wedge e^{b})\wedge R_{ab}
    +\tfrac{1}{2}g_{xy}d\phi^{x}\wedge \star d\phi^{y}
  -\tfrac{1}{4}\mathcal{M}_{MN}F^{M} \wedge \star F^{N}\right\}\,.
\end{equation}

Observe that the last term vanishes automatically when we use the twisted
self-duality constraint. However, we are only going to impose it on the
equations of motion, which are given by 

\begin{subequations}
\label{eqmoto2}
\begin{align}
  \mathbf{E}_{c}
  & =
    \imath_{c} \star \left(e^{a} \wedge e^{b}\right) \wedge R_{ab}
    +\tfrac{1}{2}g_{xy}\left(\imath_{c}d\phi^{x}\wedge\star d\phi^{y}
    +d\phi^{x}\wedge\imath_{c}\star d\phi^{y} \right)
  \nonumber \\
  & \nonumber \\
  & \hspace{.5cm}
    -\tfrac{1}{4} \mathcal{M}_{MN}\left(\imath_{c} F^{M} \wedge \star F^{N}
    +\xi^{2} F^{M} \wedge \imath_{c} \star F^{N}\right) \,,
  \\
  & \nonumber \\
  \label{eq:Ez3}
  \mathbf{E}_{z}
  & =
    - g_{zw}\left[d\star d\phi^{w}
    +\Gamma_{xy}{}^{w}d\phi^{x}\wedge \star d\phi^{y}\right]
    -\tfrac{1}{4}\partial_{z} \mathcal{M}_{MN} F^{M} \wedge \star F^{N} \,,
  \\
  & \nonumber \\
  \mathbf{E}_{M}
  & =
    \tfrac{1}{2}d \left( \mathcal{M}_{MN} \star F^{N} \right)\,.
  % \\
  % & \nonumber \\
  % \mathbf{\Theta}
  % & =
  %   - \star \left(e^{a} \wedge e^{b} \right) \wedge \delta \omega_{ab}
  %   -\tfrac{1}{2} \mathcal{M}_{MN} \delta \mathcal{A}^{M} \wedge \star F^{N}
  %   +g_{xy} \delta \phi^{x} \star d\phi^{y}  \,.
	\end{align}
\end{subequations}

Observe that scalar equation of motion has exactly the same form as the one
coming from the original action Eq.~(\ref{eq:Ez2}) and this form will not
change when we use the twisted self-duality constraint. Using this constraint
in the last equation and using Eq.~(\ref{eq:MpreservesOmega}) we see that it
takes the form Eq.~(\ref{eq:MaxwellplusBianchis}). Finally, using the
constraint in the Einstein equations brings the energy-momentum tensor of the
potentials to the form Eq.~(\ref{eq:emtensorforms}). Thus, all the original
equations of motion are recovered upon use of the twisted self-duality
constraint.

Let us now consider the dualization of the scalars.

%%%%%%%%%%%%%%%%%%%%%%%%%%%%%%%%%%%%%%%%%%%%%%%%%%%%%%%%%%%%%%%%%%%%%%
%%%%%%%%%%%%%%%%%%%%%%%%%%%%%%%%%%%%%%%%%%%%%%%%%%%%%%%%%%%%%%%%%%%%%%
%%%%%%%%%%%%%%%%%%%%%%%%%%%%%%%%%%%%%%%%%%%%%%%%%%%%%%%%%%%%%%%%%%%%%%
%%%%%%%%%%%%%%%%%%%%%%%%%%%%%%%%%%%%%%%%%%%%%%%%%%%%%%%%%%%%%%%%%%%%%%
\subsubsection{Democratic pseudoaction II: the scalars}
%%%%%%%%%%%%%%%%%%%%%%%%%%%%%%%%%%%%%%%%%%%%%%%%%%%%%%%%%%%%%%%%%%%%%% 
%%%%%%%%%%%%%%%%%%%%%%%%%%%%%%%%%%%%%%%%%%%%%%%%%%%%%%%%%%%%%%%%%%%%%%
%%%%%%%%%%%%%%%%%%%%%%%%%%%%%%%%%%%%%%%%%%%%%%%%%%%%%%%%%%%%%%%%%%%%%%
%%%%%%%%%%%%%%%%%%%%%%%%%%%%%%%%%%%%%%%%%%%%%%%%%%%%%%%%%%%%%%%%%%%%%%

Following the general prescription, we start by computing the
Noether-Gaillard-Zumino currents in the democratic theory we have just
constructed in which $F^{M}=dA^{M}$. Hitting the scalar equations of motion
with the Killing vectors $k_{A}{}^{z}$ Eq.~(\ref{eq:Ez3}) and using the
Killing vector equations and the condition
Eq.~(\ref{eq:equivarianceconditionM}) we get

\begin{equation}
  \begin{aligned}
      k_{A}{}^{z}\mathbf{E}_{z}
  & =
    - k_{A\, w}\left[d\star d\phi^{w}
    +\Gamma_{xy}{}^{w}d\phi^{x}\wedge \star d\phi^{y}\right]
    -\tfrac{1}{4}k_{A}{}^{z}\partial_{z}
    \mathcal{M}_{MN} F^{M} \wedge \star F^{N}
    \\
    & \nonumber \\
    & =
      -d\star \hat{k}_{A}
      +\tfrac{1}{2}
      T_{A}{}^{M}{}_{P}F^{P} \wedge \mathcal{M}_{MQ}\star F^{Q}
          \\
    & \nonumber \\
    & =
      -d\left[\star \hat{k}_{A}
      -\tfrac{1}{2}
      T_{A}{}^{M}{}_{P}A^{P} \wedge \mathcal{M}_{MQ}\star F^{Q}\right]
      +(-1)^{p}T_{A}{}^{M}{}_{P}A^{P} \wedge\mathbf{E}_{M}\,.
  \end{aligned}
\end{equation}

Using the twisted self-duality constraint we get a more conventional form 

\begin{equation}
      k_{A}{}^{z}\mathbf{E}_{z}
   =
      -d\left[\star \hat{k}_{A}
      -\tfrac{1}{2}
      T_{A}{}^{M}{}_{P}\Omega_{MN}A^{P} \wedge F^{N}\right]
      +(-1)^{p}T_{A}{}^{M}{}_{P}A^{P} \wedge\mathbf{E}_{M}\,,
\end{equation}

\noindent
which leads to on-shell conserved NGZ currents

\begin{equation}
  J_{A}
  \equiv
  \star \hat{k}_{A}
  -\tfrac{1}{2} T_{A}{}^{M}{}_{P}\Omega_{MN}A^{P} \wedge F^{N}\,,
\end{equation}

\noindent
and to the definition of the dual $(d-2)$-forms $C_{A}$

\begin{equation}
  dC_{A}
  \equiv
  \star \hat{k}_{A}
  -\tfrac{1}{2} T_{A}{}^{M}{}_{P}\Omega_{MN}A^{P} \wedge F^{N}\,,
\end{equation}

\noindent
and of their gauge-invariant field strengths $G_{A}$

\begin{equation}
\label{eq:dualityrelation2}
  G_{A}
  \equiv
  dC_{A}+\tfrac{1}{2} T_{A}{}^{M}{}_{P}\Omega_{MN}A^{P} \wedge F^{N}
  =
  \star \hat{k}_{A}\,.
\end{equation}

The gauge invariance of $G_{A}$ follows from that of the Killing vector and
implies that the dual $(d-2)$-form potentials $C_{A}$ and the $(p+1)$-form
potentials $A^{M}$ transform as

\begin{equation}
  \delta_{\Lambda}A^{M} = d\Lambda^{M}\,,
  \hspace{1cm}
  \delta_{\Lambda}C_{A} = d \Lambda_{A}
  -\tfrac{1}{2} T_{A}{}^{M}{}_{P}\Omega_{MN}\Lambda^{P} \wedge F^{N}\,,
\end{equation}

\noindent
where $\Lambda^{M}$ and $\Lambda_{A}$ are arbitrary $p$- and $(d-3)$-forms,
respectively.

Based on our previous results, we propose the fully democratic pseudoaction

\begin{equation}
  \label{eq:democraticII}
  \begin{aligned}
  S[e^{a},A^{M},\phi^{x},C_{A}]
   & =
  \int\left\{-\star (e^{a}\wedge e^{b})\wedge R_{ab}
    +\tfrac{1}{4}g_{xy}d\phi^{x}\wedge \star d\phi^{y}
  \right.
    \\
    & \\
    & \hspace{.5cm}
      \left.
    -\tfrac{1}{4}\mathcal{M}_{MN}F^{M} \wedge \star F^{N}
    +\tfrac{1}{4} \mathfrak{M}^{AB} G_{A} \wedge \star G_{B}
    -\tfrac{1}{2} g^{AB} G_{A} \wedge \hat{k}_{B}\right\}\,,
  \end{aligned}
\end{equation}

\noindent
whose equations of motion have to be supplemented by the twisted self-duality
constraint Eq.~(\ref{eq:twistedselfduality}) and by the duality relation
Eq.~(\ref{eq:dualityrelation2}).

The equations of motion that follow from the above action are

\begin{subequations}
\label{eqmoto3}
\begin{align}
  \mathbf{E}_{c}
  & =
    \imath_{c} \star \left(e^{a} \wedge e^{b}\right) \wedge R_{ab}
    +\tfrac{1}{2}g_{xy}\left(\imath_{c}d\phi^{x}\wedge\star d\phi^{y}
    +d\phi^{x}\wedge\imath_{c}\star d\phi^{y} \right)
  \nonumber \\
  & \nonumber \\
  & \hspace{.5cm}
    -\tfrac{1}{4} \mathcal{M}_{MN}\left(\imath_{c} F^{M} \wedge \star F^{N}
    +\xi^{2} F^{M} \wedge \imath_{c} \star F^{N}\right)
 \nonumber  \\
  & \nonumber \\
  & \hspace{.5cm}
    +\frac{1}{4} \mathfrak{M}^{AB} \left(\imath_{c} G_{A} \wedge \star G_{B}
    +G_{A} \wedge \imath_{c} \star G_{B}\right)\,,
  \\
  & \nonumber \\
  \mathbf{E}_{z}
  & =
    -\tfrac{1}{2}g_{zw}\left[d\star d\phi^{w}
    +\Gamma_{xy}{}^{w}d\phi^{x}\wedge \star d\phi^{y}\right]
    -\tfrac{1}{4}\partial_{z} \mathcal{M}_{MN} F^{M} \wedge \star F^{N}
     \nonumber  \\
  & \nonumber \\
  & \hspace{.5cm}
    +\tfrac{1}{4}\partial_{z}\mathfrak{M}^{AB}G_{A}\wedge\star G_{B}
    -\tfrac{1}{2}g^{AB}G_{A}\wedge d\phi^{x} \partial_{z}k_{B\, x}
 -\tfrac{1}{2}d\left(g^{AB}G_{A}k_{B\, z}\right)\,,
  \\
  & \nonumber \\
  \mathbf{E}_{M}
  & =
    \tfrac{1}{2}d \left( \mathcal{M}_{MN} \star F^{N} \right)
    +\tfrac{1}{2}\frac{\delta G_{A}}{A^{M}}\wedge
    \left(\mathfrak{M}^{AB}\star G_{B}-g^{AB}\hat{k}_{B}\right)\,,
  \\
  & \nonumber \\
  \mathbf{E}^{A}
  & =
    -\tfrac{1}{2}d\left(\mathfrak{M}^{AB}\star G_{B}-g^{AB}\hat{k}_{B}\right)\,.
\end{align}
\end{subequations}

Using the duality constraints and following the same steps as in previous
sections we recover the equations of motion of the original theory.

%%%%%%%%%%%%%%%%%%%%%%%%%%%%%%%%%%%%%%%%%%%%%%%%%%%%%%%%%%%%%%%%%%%%%%
%%%%%%%%%%%%%%%%%%%%%%%%%%%%%%%%%%%%%%%%%%%%%%%%%%%%%%%%%%%%%%%%%%%%%%
%%%%%%%%%%%%%%%%%%%%%%%%%%%%%%%%%%%%%%%%%%%%%%%%%%%%%%%%%%%%%%%%%%%%%%
%%%%%%%%%%%%%%%%%%%%%%%%%%%%%%%%%%%%%%%%%%%%%%%%%%%%%%%%%%%%%%%%%%%%%%
\section{A democratic pseudoaction for $\mathcal{N}=2B,d=10$ supergravity}
\label{sec-2B}
%%%%%%%%%%%%%%%%%%%%%%%%%%%%%%%%%%%%%%%%%%%%%%%%%%%%%%%%%%%%%%%%%%%%%% 
%%%%%%%%%%%%%%%%%%%%%%%%%%%%%%%%%%%%%%%%%%%%%%%%%%%%%%%%%%%%%%%%%%%%%%
%%%%%%%%%%%%%%%%%%%%%%%%%%%%%%%%%%%%%%%%%%%%%%%%%%%%%%%%%%%%%%%%%%%%%%
%%%%%%%%%%%%%%%%%%%%%%%%%%%%%%%%%%%%%%%%%%%%%%%%%%%%%%%%%%%%%%%%%%%%%%

The bosonic fields of $\mathcal{N}=2B,d=10$ supergravity are the
(Einstein-frame) Zehnbein 1-form $e^{a}$, a SL$(2,\mathbb{R})$ doublet of
2-forms $\mathcal{B}_{i}$, $i=1,2$, 4-form $\mathcal{D}$, which is a
SL$(2,\mathbb{R})$ singlet and whose 5-form field strength is self-dual and
the complex scalar $\tau$ that parametrizes a SL$(2,\mathbb{R})/$SO$(2)$
coset.  The relation between these fields and those of the effective action of
the type~IIB superstring (string-frame Zehnbein $e^{a}_{\rm s}$, NSNS and RR
2-forms $B,C^{(2)}$, RR 4-form $C^{(4)}$ and dilaton $\varphi$ and RR 0-form
$C^{(0)}$) is \cite{Ortin:2015hya}\footnote{Observe that the RR 4-form
  $C^{(4)}$ is not an SL$(2,\mathbb{R})$ and, as a matter of fact, transforms
  in a complicated way under those transformations. On the other hand, in the
  rescaling between the string- and Einstein-frame metrics it is very
  important to take into account the effect of the constant value of the
  dilaton at infinity, in order to preserve the standard normalization of the
  metric at infinity \cite{Maldacena:1996ky}. The relation between the string
  and the so-called \textit{modified Einstein frame} should, then, be
  $e^{a} = e^{-(\varphi-\varphi_{\infty})/4}e^{a}_{\rm s}$. This leads to
  the occurrence of factors of powers of $e^{\varphi_{\infty}/4}$ in in
  different terms of the action that have to be removed by absorbing them in
  redefinitions of the rest of the fields. We will not study these
  redefinitions here because they are not relevant to the construction of the
  democratic pseudoaction.}

\begin{equation}
  \begin{aligned}
    e^{a}
    & = e^{-\varphi/4}e^{a}_{\rm s}\,,
    \\
    & \\
    \tau
    & = C^{(0)}+i e^{-\varphi}\,,
    \\
    & \\
    \left(\mathcal{B}_{i}\right)
    & =
      \left(
      \begin{array}{c}
C^{(2)} \\ B \\
      \end{array}
      \right)\,,
    \\
    & \\
    \mathcal{D}
    & =
      C^{(4)}-\tfrac{1}{2}B\wedge C^{(2)}\,.    
  \end{aligned}
\end{equation}

The self-duality of the 5-form field strength forbids the existence of a
covariant action free of auxiliary fields and one, if one does not want to
deal with auxiliary fields, one must necessarily work with equations of
motion. The equations of motion of this theory were first found in
Ref.~\cite{Schwarz:1983qr} in the Einstein frame and using a SU$(1,1)/$U$(1)$
formulation of the coset space parametrized by the scalar fields. A
pseudoaction which had to be supplemented by the self-duality constraint was
first constructed in Ref.~\cite{Bergshoeff:1995sq}. A pseudoaction containing
the duals of the 1- and 3-form RR field strengths and some higher RR field
strengths was constructed in Ref.~\cite{Bergshoeff:2001pv}. While this action
was ``RR-democratic'' it was certainly not ``NSNS-democratic''. Furthermore,
this incomplete democratization breaks the manifest SL$(2,\mathbb{R})$
symmetry of the theory. Our goal in this section is to improve on those
results constructing a manifestly SL$(2,\mathbb{R})$-invariant and fully
democratic pseudoaction using the results of the previous sections.

The transformation of the scalars under SL$(2,\mathbb{R})$ can be conveniently
described through the symmetric SL$(2,\mathbb{R})$ matrix

\begin{equation}
  \left(\mathcal{M}_{ij} \right)
  \equiv
  \frac{1}{\Im\mathrm{m}\, \tau}
  \left(
    \begin{array}{lr}
      |\tau|^{2} & \Re\mathrm{e}\,\tau \\
        & \\
       \Re\mathrm{e}\,\tau & 1 \\
    \end{array}
  \right)\,,
\end{equation}

\noindent
whose inverse is

\begin{equation}
  \left(\mathcal{M}^{ij} \right)
  \equiv
  \frac{1}{\Im\mathrm{m}\, \tau}
  \left(
    \begin{array}{lr}
      1 & -\Re\mathrm{e}\,\tau \\
        & \\
       -\Re\mathrm{e}\,\tau & |\tau|^{2} \\
    \end{array}
  \right)\,.  
\end{equation}

\noindent
If we act with the SL$(2,\mathbb{R})$ transformation matrix

\begin{equation}
  \left(\mathcal{S}^{-1\, i}{}_{j}\right)
  \equiv
  \left(
    \begin{array}{lr}
      \alpha & \gamma \\
       \beta & \delta \\
    \end{array}
  \right)\,,
  \hspace{1cm}
  \alpha\delta -\beta\gamma = +1\,,
\end{equation}

\noindent
on objects with indices

\begin{equation}
  \mathcal{B}_{i}'
  =
  \mathcal{B}_{j} \mathcal{S}^{-1\, j}{}_{i}\,,
  \hspace{1cm}
  \mathcal{M}_{ij}'
  =
  \mathcal{M}_{kl}\mathcal{S}^{-1\, k}{}_{i}\mathcal{S}^{-1\, l}{}_{j}\,,
  \hspace{1cm}
  \mathcal{M}^{ij\, \prime}
  =
  \mathcal{S}^{i}{}_{k}\mathcal{S}^{j}{}_{l}\mathcal{M}^{kl}\,,
\end{equation}

\noindent
then $\tau$ transforms as

\begin{equation}
  \tau'
  =
  \frac{\alpha\tau +\beta}{\gamma \tau+\delta}\,.
\end{equation}

The field strength of the doublet of 2-forms is the doublet of 3-forms

\begin{equation}
  \mathcal{H}_{i}
  \equiv
  d\mathcal{B}_{i}\,,
\end{equation}

\noindent
while the field strength of the 4-form is the SL$(2,\mathbb{R})$-invariant
5-form

\begin{equation}
  \mathcal{F}
  \equiv
  d\mathcal{D}
  -\tfrac{1}{2} \varepsilon^{ij}\mathcal{B}_{i}\wedge \mathcal{H}_{j}\,.
\end{equation}

The doublet of 3-form field strengths $\mathcal{H}_{i}$ and the 5-form
$\mathcal{F}$ are invariant under the gauge transformations

\begin{equation}
  \label{eq:gaugetransformations2and4}
  \delta_{\Lambda} \mathcal{B}_{i}
  =
  d\Lambda_{i}\,,
  \hspace{1cm}
  \delta_{\Lambda} \mathcal{D}
  =
  d\Lambda +\tfrac{1}{2} \varepsilon^{ij}\Lambda_{i}\wedge \mathcal{H}_{j}\,,
\end{equation}

\noindent
where $\Lambda$ and $\Lambda_{i}$ are, respectively, a 4-form and a doublet of
1-forms.

This field strength is constrained to be self-dual\footnote{This constraint
  is, actually, the equation of motion, \textit{sensu stricto}.}

\begin{equation}
  \label{eq:F=starF}
  \mathcal{F}
  =
  \star \mathcal{F}\,,  
\end{equation}

\noindent
and this condition relates the Bianchi identity

\begin{equation}
  d\mathcal{F}
  +\tfrac{1}{2}\varepsilon^{ij}\mathcal{H}_{i}\wedge \mathcal{H}_{j}
  =
  0\,,  
\end{equation}

\noindent
to the equation of motion \cite{Schwarz:1983qr}
 
\begin{equation}
  d\star\mathcal{F}
  +\tfrac{1}{2}\varepsilon^{ij} \mathcal{H}_{i}\wedge \mathcal{H}_{j}
  =
  0\,,  
\end{equation}

Since it also implies that $\mathcal{F}\wedge \star \mathcal{F}=0$, this
constraint makes it impossible to write a covariant action for the
theory. Ignoring it, one can write a pseudoaction which leads to the above
equation of motion for $\mathcal{D}$ (and to the right equations of motion for
the rest of the fields \cite{Bergshoeff:1995sq}. This pseudoaction can be
written in the manifestly SL$(2,\mathbb{R})$-invariant form

\begin{equation}
  \begin{aligned}
    S[e^{a},\tau,\mathcal{B}_{i},D]
    & =
       \int \left\{
      -\star(e^{a}\wedge e^{b})\wedge R_{ab}
      +\frac{d\tau\wedge \star d\bar{\tau}}{2(\Im{\rm m}\, \tau)^{2}}
      +\tfrac{1}{2}\mathcal{M}^{ij}\mathcal{H}_{i}\wedge \star \mathcal{H}_{j}
      \right.
    \\
    & \\
    & \hspace{.5cm}
      \left.
      +\tfrac{1}{4}\mathcal{F}\wedge \star \mathcal{F}
      -\tfrac{1}{4} \varepsilon^{ij}
      \mathcal{D} \wedge \mathcal{H}_{i}\wedge \mathcal{H}_{j}
      \right\}\,,
  \end{aligned}
\end{equation}

\noindent
and the equations of motion have to be supplemented by the self-duality
constraint Eq.~(\ref{eq:F=starF}).

Our goal is to generalize this pseudoaction to include the 8-form duals of the
scalar fields (a SL$(2,\mathbb{R})$ triplet) as well as the 6-form duals of
the 2-form fields (a dual SL$(2,\mathbb{R})$ doublet). We start by writing all
the equations of motion. It is convenient to define the real scalars
$\{\phi^{x}\}$ by $\tau = C^{(0)}+ie^{-\varphi} \equiv \phi^{1}+i\phi^{2}$, so
that

\begin{equation}
\frac{d\tau\wedge \star d\bar{\tau}}{2(\Im{\rm m}\, \tau)^{2}}
=
\tfrac{1}{2}g_{xy}d\phi^{x}\wedge \star d\phi^{y}\,,
\hspace{1cm}
(g_{xy})
=
\frac{1}{(\phi^{2})^{2}}
\left(
    \begin{array}{lr}
      1 & 0 \\
       0 & 1 \\
    \end{array}
  \right)\,.
\end{equation}

Then, the equations of motion take the form

\begin{subequations}
  \begin{align}
    \mathbf{E}_{a}
    & =
     \imath_{a}\star (e^{c}\wedge e^{d})\wedge R_{cd}
      +\tfrac{1}{2}g_{xy}
    \left(\imath_{a}d\phi^{x}\wedge \star d\phi^{y}
      +d\phi^{x}\wedge \imath_{a}\star d\phi^{y}\right)
      \nonumber \\
    & \nonumber \\
    & \hspace{.5cm}
      +\tfrac{1}{2}\mathcal{M}^{ij}
      \left(\imath_{a}\mathcal{H}_{i}\wedge \star \mathcal{H}_{j}
      +\mathcal{H}_{i}\wedge \imath_{a}\star \mathcal{H}_{j}\right)
      +\tfrac{1}{4}\left(\imath_{a}\mathcal{F}\wedge \star \mathcal{F}
      +\mathcal{F}\wedge \imath_{a}\star \mathcal{F}\right)\,,
    \\
    & \nonumber \\
        \mathbf{E}_{x}
    & =
      -g_{xy}\left[d\star d\phi^{y}
      +\Gamma_{zw}{}^{y}d\phi^{z}\wedge \star d\phi^{w}\right]
      +\tfrac{1}{2}\partial_{x}\mathcal{M}^{ij}
      \mathcal{H}_{i}\wedge \star \mathcal{H}_{j}\,,
    \\
    & \nonumber \\
        \mathbf{E}^{i}
    & =
      -\left\{d\left(\mathcal{M}^{ij}\star\mathcal{H}_{j}\right)
      +\varepsilon^{ij}\mathcal{H}_{j}\wedge \mathcal{F}\right\}
      +\tfrac{1}{2}\varepsilon^{ij}\mathcal{H}_{j}\wedge\left(
      \mathcal{F}-\star \mathcal{F} \right)
      +\tfrac{1}{2}\varepsilon^{ij}\mathcal{B}_{j}\wedge \mathbf{E}\,,
    \\
    & \nonumber \\
        \mathbf{E}
    & =
      -\tfrac{1}{2}
      \left\{ d\star\mathcal{F}
      +\tfrac{1}{2}\varepsilon^{ij}
      \mathcal{H}_{i}\wedge \mathcal{H}_{j}\right\}\,.
  \end{align}
\end{subequations}

The Einstein equations and the equations of motion of the 2-forms and 4-form
simplify when the self-duality constraint is imposed. In particular, since the
equation of motion of the 4-form becomes the Bianchi identity, it is
automatically solved. The remaining non-trivial equations of motion are

\begin{subequations}
  \begin{align}
    \mathbf{E}_{a}
    & =
     \imath_{a}\star (e^{c}\wedge e^{d})\wedge R_{cd}
      +\tfrac{1}{2}g_{xy}
    \left(\imath_{a}d\phi^{x}\wedge \star d\phi^{y}
      +d\phi^{x}\wedge \imath_{a}\star d\phi^{y}\right)
      \nonumber \\
    & \nonumber \\
    & \hspace{.5cm}
      +\tfrac{1}{2}\mathcal{M}^{ij}
      \left(\imath_{a}\mathcal{H}_{i}\wedge \star \mathcal{H}_{j}
      +\mathcal{H}_{i}\wedge \imath_{a}\star \mathcal{H}_{j}\right)
      +\tfrac{1}{2}\imath_{a}\mathcal{F}\wedge \mathcal{F}\,,
    \\
    & \nonumber \\
        \mathbf{E}_{x}
    & =
      -g_{xy}\left[d\star d\phi^{y}
      +\Gamma_{zw}{}^{y}d\phi^{z}\wedge \star d\phi^{w}\right]
      +\tfrac{1}{2}\partial_{x}\mathcal{M}^{ij}
      \mathcal{H}_{i}\wedge \star \mathcal{H}_{j}\,,
    \\
    & \nonumber \\
        \mathbf{E}^{i}
    & =
      -d\left(\mathcal{M}^{ij}\star\mathcal{H}_{j}\right)
      -\varepsilon^{ij}\mathcal{H}_{j}\wedge \mathcal{F}\,,
  \end{align}
\end{subequations}

\noindent
together with the self-duality constraint Eq.~(\ref{eq:F=starF}).

%%%%%%%%%%%%%%%%%%%%%%%%%%%%%%%%%%%%%%%%%%%%%%%%%%%%%%%%%%%%%%%%%%%%%%
%%%%%%%%%%%%%%%%%%%%%%%%%%%%%%%%%%%%%%%%%%%%%%%%%%%%%%%%%%%%%%%%%%%%%%
%%%%%%%%%%%%%%%%%%%%%%%%%%%%%%%%%%%%%%%%%%%%%%%%%%%%%%%%%%%%%%%%%%%%%%
%%%%%%%%%%%%%%%%%%%%%%%%%%%%%%%%%%%%%%%%%%%%%%%%%%%%%%%%%%%%%%%%%%%%%%
\subsection{Dualization of the 2-forms}
%%%%%%%%%%%%%%%%%%%%%%%%%%%%%%%%%%%%%%%%%%%%%%%%%%%%%%%%%%%%%%%%%%%%%%
%%%%%%%%%%%%%%%%%%%%%%%%%%%%%%%%%%%%%%%%%%%%%%%%%%%%%%%%%%%%%%%%%%%%%%
%%%%%%%%%%%%%%%%%%%%%%%%%%%%%%%%%%%%%%%%%%%%%%%%%%%%%%%%%%%%%%%%%%%%%%
%%%%%%%%%%%%%%%%%%%%%%%%%%%%%%%%%%%%%%%%%%%%%%%%%%%%%%%%%%%%%%%%%%%%%%

In order to construct the democratic pseudoaction, we start by considering the
dualization of the doublet of 2-forms $\mathcal{B}_{i}$ into a doublet of
6-forms that we will denote by $\mathcal{B}^{i}$.

The equations of motion of the doublet of 2-forms can be written as a total
derivative:

\begin{equation}
        \mathbf{E}^{i}
    =
    -d\left[\mathcal{M}^{ij}\star\mathcal{H}_{j}
      +\varepsilon^{ij}\mathcal{B}_{j}\wedge
      \left(d\mathcal{D}
        -\tfrac{1}{6}\varepsilon^{kl}\mathcal{B}_{k}\wedge
        \mathcal{H}_{l}\right)
    \right]\,,  
\end{equation}

\noindent
and, therefore, they can be locally solved by identifying the expression in
square brackets with $d\mathcal{B}^{i}$. Then,

\begin{equation}
  \mathcal{M}^{ij}\star\mathcal{H}_{j}
  =
  d\mathcal{B}^{i} -\varepsilon^{ij}\mathcal{B}_{j}\wedge
      \left(d\mathcal{D}
        -\tfrac{1}{6}\varepsilon^{kl}\mathcal{B}_{k}\wedge
        \mathcal{H}_{l}\right)
      \equiv
      \mathcal{H}^{i}\,,
\end{equation}

\noindent
where $ \mathcal{H}^{i}$ is the SL$(2,\mathbb{R})$ doublet of 7-form field
strengths, invariant under the gauge transformations
Eqs.~(\ref{eq:gaugetransformations2and4}) if the 6-forms transform according
to

\begin{equation}
  \label{eq:gaugetransformations6}
  \delta_{\Lambda}\mathcal{B}^{i}
  =
  d\Lambda^{i}
  +\varepsilon^{ij}\Lambda_{j}\wedge d\mathcal{D}
  -\tfrac{1}{6}\varepsilon^{ij}\varepsilon^{lk}\mathcal{B}_{j}\wedge
  \mathcal{B}_{l} \wedge d\Lambda_{k}\,,
\end{equation}

\noindent
where  $\Lambda^{i}$ is a doublet of 5-forms.

The Bianchi identity of the 3-form field strengths

\begin{equation}
d \mathcal{H}_{i}= 0\,,
\end{equation}

\noindent
becomes the equation of motion of the 6-forms upon use of the duality relation

\begin{equation}
  \label{eq:2-6duality}
  \mathcal{H}_{i}
  =
  \mathcal{M}_{ij}\star \mathcal{H}^{j}\,,
\end{equation}

\noindent
that is

\begin{equation}
 \label{eq:eom6forms}
d\left( \mathcal{M}_{ij}\star \mathcal{H}^{j}\right)= 0\,.
\end{equation}

The dual 6-forms that we have just defined can easily be included in a
semi-democratic pseudoaction\footnote{Observe that the sign of the
  Chern-Simons term has been reversed.}

\begin{equation}
\label{eq:semidemocraticaction}
  \begin{aligned}
    S_{\rm SemiDem}[e^{a},\tau,\mathcal{B}_{i},D,\mathcal{B}^{i}]
    & =
       \int \left\{
      -\star(e^{a}\wedge e^{b})\wedge R_{ab}
      +\tfrac{1}{2}g_{xy}d\phi^{x}\wedge \star d\phi^{y}
      +\tfrac{1}{4}\mathcal{M}^{ij}\mathcal{H}_{i}\wedge \star \mathcal{H}_{j}
      \right.
    \\
    & \\
    & \hspace{.5cm}
      \left.
      +\tfrac{1}{4}\mathcal{F}\wedge \star \mathcal{F}
      +\tfrac{1}{4}\mathcal{M}_{ij}\mathcal{H}^{i}\wedge \star \mathcal{H}^{j}
      +\tfrac{1}{4} \varepsilon^{ij}
      \mathcal{D} \wedge \mathcal{H}_{i}\wedge \mathcal{H}_{j}
      \right\}\,.
  \end{aligned}
\end{equation}

The equations of motion that follow from this pseudoaction are

\begin{subequations}
\label{eq:semidemocraticeoms}
  \begin{align}
    \mathbf{E}_{a}
    & =
     \imath_{a}\star (e^{c}\wedge e^{d})\wedge R_{cd}
      +\tfrac{1}{2}g_{xy}
    \left(\imath_{a}d\phi^{x}\wedge \star d\phi^{y}
      +d\phi^{x}\wedge \imath_{a}\star d\phi^{y}\right)
      \nonumber \\
    & \nonumber \\
    & \hspace{.5cm}
      +\tfrac{1}{4}\mathcal{M}^{ij}
      \left(\imath_{a}\mathcal{H}_{i}\wedge \star \mathcal{H}_{j}
      +\mathcal{H}_{i}\wedge \imath_{a}\star \mathcal{H}_{j}\right)
      +\tfrac{1}{4}\left(\imath_{a}\mathcal{F}\wedge \star \mathcal{F}
      +\mathcal{F}\wedge \imath_{a}\star \mathcal{F}\right)
      \nonumber \\
    & \nonumber \\
    & \hspace{.5cm}
      +\tfrac{1}{4}\mathcal{M}_{ij}
      \left(\imath_{a}\mathcal{H}^{i}\wedge \star \mathcal{H}^{j}
      +\mathcal{H}^{i}\wedge \imath_{a}\star \mathcal{H}^{j}\right)\,,
    \\
    & \nonumber \\
        \mathbf{E}_{x}
    & =
      -g_{xy}\left[d\star d\phi^{y}
      +\Gamma_{zw}{}^{y}d\phi^{z}\wedge \star d\phi^{w}\right]
      +\tfrac{1}{4}\partial_{x}\mathcal{M}^{ij}
      \mathcal{H}_{i}\wedge \star \mathcal{H}_{j}
      +\tfrac{1}{4}\partial_{x}\mathcal{M}_{ij}
      \mathcal{H}^{i}\wedge \star \mathcal{H}^{j}\,,
    \\
    & \nonumber \\
        \mathbf{E}^{i}
    & =
      -\tfrac{1}{2}\left\{
      d\left(\mathcal{M}^{ij}\star\mathcal{H}_{j}\right)
      +\varepsilon^{ij}\mathcal{H}_{j}\wedge \mathcal{F}
      \right\}
      +\tfrac{1}{2}\varepsilon^{ij}\mathcal{H}_{j}\wedge
      (\star\mathcal{F}-\mathcal{F})
      \nonumber \\
    & \nonumber \\
    & \hspace{.5cm}
      +\tfrac{1}{2}\varepsilon^{ij}
            \left(\mathcal{H}_{j}-\mathcal{M}_{jk}\star\mathcal{H}^{k}\right)\wedge \mathcal{F}
      +\tfrac{1}{2}\varepsilon^{ij}\mathcal{B}_{j}\wedge \mathbf{E}
      -\tfrac{2}{3}\varepsilon^{ij}\varepsilon^{kl}\mathcal{B}_{j}\wedge
      \mathcal{B}_{k}\wedge \mathbf{E}_{l}\,,
    \\
    & \nonumber \\
        \mathbf{E}
    & =
      -\tfrac{1}{2}
      \left\{ d\star\mathcal{F}
      -\tfrac{1}{2}\varepsilon^{ij} \mathcal{H}_{i}\wedge \left(
      \mathcal{H}_{j}-2\mathcal{M}_{jk}\star \mathcal{H}^{k}\right)\right\}
      +\varepsilon^{ij}\mathcal{B}_{i}\wedge \mathbf{E}_{j}\,,
    \\
    & \nonumber \\
        \mathbf{E}_{i}
    & =
      -\tfrac{1}{2}d\left(\mathcal{M}_{ij}\star\mathcal{H}^{j}\right)\,.
  \end{align}
\end{subequations}

Upon use of the duality relations Eq.~(\ref{eq:2-6duality}) the equations of
motion of the 6-forms $\mathbf{E}_{i}$ become the Bianchi identities of the
3-forms and are automatically solved. Furthermore, using the same duality
relations plus the self-duality constraint Eq.~(\ref{eq:F=starF}) the last two
lines and the third term of the first line of the equations of motion of the
2-forms $\mathbf{E}^{i}$ vanish and what remains becomes, up to a factor of
$1/2$, the original equations of motion of the 2-forms. The rest of the
equations of motion are trivially recovered.

%%%%%%%%%%%%%%%%%%%%%%%%%%%%%%%%%%%%%%%%%%%%%%%%%%%%%%%%%%%%%%%%%%%%%%
%%%%%%%%%%%%%%%%%%%%%%%%%%%%%%%%%%%%%%%%%%%%%%%%%%%%%%%%%%%%%%%%%%%%%%
%%%%%%%%%%%%%%%%%%%%%%%%%%%%%%%%%%%%%%%%%%%%%%%%%%%%%%%%%%%%%%%%%%%%%%
%%%%%%%%%%%%%%%%%%%%%%%%%%%%%%%%%%%%%%%%%%%%%%%%%%%%%%%%%%%%%%%%%%%%%%
\subsection{Dualization of the  scalars}
%%%%%%%%%%%%%%%%%%%%%%%%%%%%%%%%%%%%%%%%%%%%%%%%%%%%%%%%%%%%%%%%%%%%%%
%%%%%%%%%%%%%%%%%%%%%%%%%%%%%%%%%%%%%%%%%%%%%%%%%%%%%%%%%%%%%%%%%%%%%%
%%%%%%%%%%%%%%%%%%%%%%%%%%%%%%%%%%%%%%%%%%%%%%%%%%%%%%%%%%%%%%%%%%%%%%
%%%%%%%%%%%%%%%%%%%%%%%%%%%%%%%%%%%%%%%%%%%%%%%%%%%%%%%%%%%%%%%%%%%%%%

The next step towards the construction of the democratic pseudoaction is the
dualization of the scalars, which in this theory parametrize the symmetric
Riemannian manifold SL$(2,\mathbb{R})/$SO$(2)$. This means that we can use the
general procedure described in Sections~\ref{sec-sigmamodelsymmetric} and
~\ref{sec-sigmamodelsymmetriccoupledtop+1forms}.

Under the three independent infinitesimal SL$(2,\mathbb{R})$ transformations
labeled by $A$, the fields that we have introduced so far transform according
to

\begin{equation}
  \delta_{A}\phi^{x}
  =
  k_{A}{}^{x}(\phi)\,,
  \hspace{1cm}
  \delta_{A} \mathcal{B}_{i}
  =
  -\mathcal{B}_{j}\left(T_{A}\right)^{j}{}_{i}\,,
  \hspace{1cm}
  \delta_{A} \mathcal{B}^{i}
  =
  \left(T_{A}\right)^{i}{}_{j}\mathcal{B}_{j}\,,
\end{equation}

\noindent
while the kinetic matrix $\mathcal{M}_{ij}$ and its inverse $\mathcal{M}^{ij}$
satisfy

\begin{equation}
  k_{A}{}^{x}\partial_{x}\mathcal{M}_{ij}
  =
  -2\mathcal{M}_{k(j}\left(T_{A}\right)^{k}{}_{i)}\,,
  \hspace{1cm}
  k_{A}{}^{x}\partial_{x}\mathcal{M}^{ij}
  =
  2\left(T_{A}\right)^{(i}{}_{k}\mathcal{M}^{j)k}\,.
\end{equation}

Observe that due to the fact that $\varepsilon^{ij}$ and $\varepsilon_{ij}$
are SL$(2,\mathbb{R})$-invariant tensors, the matrices $T_{A}$ satisfy

\begin{equation}
  \left(T_{A}\right)^{[i}{}_{k}\varepsilon^{j]k}
  =
  \left(T_{A}\right)^{k}{}_{[i}\varepsilon_{j]k}
  =
  0\,.
\end{equation}

Then, we can obtain the Noether-Gaillard-Zumino 9-form currents $J_{A}$ using
the Killing vector equation, the duality relations and the equations of motion
of the 2- and 6-forms and these properties, obtaining

\begin{equation}
  \begin{aligned}
    k_{A}{}^{x}\mathbf{E}_{x}
    & =
      -d\left[\star\hat{k}_{A}
      -\tfrac{1}{2} \left(T_{A}\right)^{i}{}_{k}
      \mathcal{B}_{i}\wedge \mathcal{H}^{k}
      +\tfrac{1}{2}\left(T_{A}\right)^{k}{}_{i}
        \mathcal{B}^{i}\wedge \mathcal{H}_{k}
        \right.
    \\
    & \\
    & \hspace{.5cm}
      \left.
      +\tfrac{1}{24}\left(T_{A}\right)^{i}{}_{k}\varepsilon^{kj} \varepsilon^{mn}
      \mathcal{B}_{i}\wedge \mathcal{B}_{j}\wedge
      \mathcal{B}_{m} \wedge \mathcal{H}_{n}\right]\,.
    % \\
    % & \\
    % & \hspace{.5cm}
    %   +\left(T_{A}\right)^{i}{}_{k}\mathcal{B}_{i}\wedge \mathbf{E}^{k}
    %   -\tfrac{1}{2}\left(T_{A}\right)^{k}{}_{i}
    %   \mathcal{B}^{i}\wedge \mathbf{E}_{k}  
  \end{aligned}
\end{equation}

This expression vanishes on-shell, and we can solve it locally by introducing
a SL$(2,\mathbb{R})$ triplet of 8-forms $C_{A}$ whose exterior derivative
equals the expression in brackets. Since the Killing vectors are gauge
invariant, we can define the following gauge-invariant triplet of 9-form fields
strengths

\begin{equation}
  \begin{aligned}
\star \hat{k}_{A}
  & =
    dC_{A}
      +\tfrac{1}{2} \left(T_{A}\right)^{i}{}_{k}
      \mathcal{B}_{i}\wedge \mathcal{H}^{k}
      -\tfrac{1}{2}\left(T_{A}\right)^{k}{}_{i}
        \mathcal{B}^{i}\wedge \mathcal{H}_{k}
    \\
    & \\
    & \hspace{.5cm}
      -\tfrac{1}{24}\left(T_{A}\right)^{i}{}_{k}\varepsilon^{kj} \varepsilon^{mn}
      \mathcal{B}_{i}\wedge \mathcal{B}_{j}\wedge
      \mathcal{B}_{m} \wedge \mathcal{H}_{n}
    \\
    & \\
    & \equiv
      G_{A}\,.
  \end{aligned}
\end{equation}

$G_{A}$ is invariant under the gauge transformations in
Eqs.~(\ref{eq:gaugetransformations2and4}) and (\ref{eq:gaugetransformations6})
if the 8-forms transform as 

\begin{equation}
  \delta C_{A}
  =
  d\Lambda_{A}
  -\tfrac{1}{2}\left(T_{A}\right)^{i}{}_{k}\left\{
    \Lambda^{k}\wedge\mathcal{H}_{i} -\Lambda_{i}\wedge \mathcal{H}^{k}
    +\tfrac{1}{4}\varepsilon^{kl}\varepsilon^{mn}
    \mathcal{B}_{i}\wedge \mathcal{B}_{l}\wedge \Lambda_{m}
    \wedge\mathcal{H}_{n}\right\}\,.
\end{equation}

Having defined the 8-form duals of the scalars and using our previous
experience, we propose the following manifestly gauge- and SL$(2,\mathbb{R})$
fully democratic pseudoaction

\begin{equation}
\label{eq:democraticaction}
  \begin{aligned}
    S_{\rm Dem}[e^{a},\tau,\mathcal{B}_{i},D,\mathcal{B}^{i},C_{A}]
    & =
      \int \left\{
      -\star(e^{a}\wedge e^{b})\wedge R_{ab}
      +\tfrac{1}{4}g_{xy}d\phi^{x}\wedge \star d\phi^{y}
      +\tfrac{1}{4}\mathcal{M}^{ij}\mathcal{H}_{i}\wedge \star \mathcal{H}_{j}
      \right.
    \\
    & \\
    & \hspace{.5cm}
      +\tfrac{1}{4}\mathcal{F}\wedge \star \mathcal{F}
      +\tfrac{1}{4}\mathcal{M}_{ij}\mathcal{H}^{i}\wedge \star \mathcal{H}^{j}
      +\tfrac{1}{4}\mathfrak{M}^{AB}G_{A}\wedge \star G_{B}
          \\
    & \\
    & \hspace{.5cm}
      \left.
      -\tfrac{1}{2}g^{AB}G_{A}\wedge \star \hat{k}_{B}
      +\tfrac{1}{4} \varepsilon^{ij}
      \mathcal{D} \wedge \mathcal{H}_{i}\wedge \mathcal{H}_{j}
      \right\}\,.
  \end{aligned}
\end{equation}

\noindent
The explicit form of this action depends on the particular choice of Killing
vector basis. A convenient choice is

\begin{equation}
  k_{1} = C^{(0)} \partial_{C} - \partial_{\varphi} \,,
  \hspace{1cm}
  k_{2} = (e^{-2\varphi} - C^{(0)}{}^{2})\partial_{C} + 2 C^{(0)}
  \partial_{\varphi}\,,
  \hspace{1cm}
  k_{3} = \partial_{C} \,.
\end{equation}

\noindent
Since, in this basis, $k_{3}$ generates the constant shifts of the RR 0-form
$C^{(0)}$, we can identify the 8-form $C_{3}$ with the RR 8-form $C^{(8)}$.

The Lie brackets of these vectors are 

\begin{equation}
  [k_{1},k_{3}] = - k_{3} \,,
  \hspace{1cm}
  [k_{2}, k_{3}] = 2 k_{1} \,,
  \hspace{1cm}
  [k_{1}, k_{2}] = k_{2} \,,
\end{equation}

\noindent
which leads to the  Killing metric $K_{AB}$

\begin{equation}
  K_{AB}
  =
  f_{AC}{}^{D} f_{BD}{}^{C}
  =
  \begin{pmatrix}
	2 & 0  & 0 \\
	0 &  0 & 4  \\
	0 & 4   &  0
	\end{pmatrix}\,.
\end{equation}

In our conventions the metric $g_{AB}$ used to construct the $\sigma$-model
metric is then related with the Killing metric by

\begin{equation}
  g_{AB}
  =
  \tfrac{1}{2} K_{AB}
  =
  \begin{pmatrix}
		1 & 0  & 0 \\
		0 &  0 & 2 \\
		0 & 2   &  0
	\end{pmatrix} \,.
\end{equation}

Then the matrix $\mathfrak{M}^{AB}$ defined in Eq.~(\ref{eq:mathfrakMABdef})
is given by

\begin{equation}
\left( \mathfrak{M}^{AB}\right)
 =
            \begin{pmatrix}
              1 + e^{2\varphi}C^{(0)}{}^{2}
              &
                \tfrac{1}{2}e^{2\varphi}C^{(0)}
              &
               -\tfrac{1}{2}C^{(0)}\left[1+e^{2\varphi}C^{(0)\, 2}\right]
              \\
              & & \\
              \tfrac{1}{2}e^{2\varphi}C^{(0)}
              &
                \frac{1}{4}e^{2\varphi}
              &
                \frac{1}{4}\left[1-e^{2\varphi}C^{(0)\, 2}\right]
              \\
              & \\
              -\tfrac{1}{2}C^{(0)}\left[1 + e^{2\varphi}C^{(0)\,2 }\right]
              &
                \frac{1}{4}\left[1-e^{2\varphi}C^{(0)\, 2}\right]
              &
                \frac{1}{4}e^{-2\varphi}\left[1 + e^{2\varphi}C^{(0)\, 2}\right]^{2} 
	\end{pmatrix} \,.
\end{equation}

We have explicitly checked that it satisfies the essential property
Eq.~(\ref{eq:Mk=gk}).

The equations of motion that follow from the democratic pseudoaction
Eq.~(\ref{eq:democraticaction}) are

\begin{subequations}
\label{eq:democraticeoms}
  \begin{align}
    \mathbf{E}_{a}
    & =
     \imath_{a}\star (e^{c}\wedge e^{d})\wedge R_{cd}
      +\tfrac{1}{4}g_{xy}
    \left(\imath_{a}d\phi^{x}\wedge \star d\phi^{y}
      +d\phi^{x}\wedge \imath_{a}\star d\phi^{y}\right)
      \nonumber \\
    & \nonumber \\
    & \hspace{.5cm}
      +\tfrac{1}{4}\mathcal{M}^{ij}
      \left(\imath_{a}\mathcal{H}_{i}\wedge \star \mathcal{H}_{j}
      +\mathcal{H}_{i}\wedge \imath_{a}\star \mathcal{H}_{j}\right)
      +\tfrac{1}{4}\left(\imath_{a}\mathcal{F}\wedge \star \mathcal{F}
      +\mathcal{F}\wedge \imath_{a}\star \mathcal{F}\right)
      \nonumber \\
    & \nonumber \\
    & \hspace{.5cm}
      +\tfrac{1}{4}\mathcal{M}_{ij}
      \left(\imath_{a}\mathcal{H}^{i}\wedge \star \mathcal{H}^{j}
      +\mathcal{H}^{i}\wedge \imath_{a}\star \mathcal{H}^{j}\right)
      +\tfrac{1}{4}\mathfrak{M}^{AB}
      \left(\imath_{a}G_{A}\wedge \star G_{B}
      +G_{A} \imath_{a}\star G_{B}\right)\,,
    \\
    & \nonumber \\
        \mathbf{E}_{x}
    & =
      -g_{xy}\left[d\star d\phi^{y}
      +\Gamma_{zw}{}^{y}d\phi^{z}\wedge \star d\phi^{w}\right]
      +\tfrac{1}{4}\partial_{x}\mathcal{M}^{ij}
      \mathcal{H}_{i}\wedge \star \mathcal{H}_{j}
      +\tfrac{1}{4}\partial_{x}\mathcal{M}_{ij}
      \mathcal{H}^{i}\wedge \star \mathcal{H}^{j}
      \nonumber \\
    & \nonumber \\
    & \hspace{.5cm}
      +\tfrac{1}{2}g^{AC}\partial_{x} k_{C\,y} G_{A} \wedge
      \left[g^{BD}k_{D}{}^{y} \star G_{B} - d\phi^{y}\right]\,,
    \\
    & \nonumber \\
        \mathbf{E}^{i}
    & =
      -\tfrac{1}{2}\left\{
      d\left(\mathcal{M}^{ij}\star\mathcal{H}_{j}\right)
      +\varepsilon^{ij}\mathcal{H}_{j}\wedge \mathcal{F}
      \right\}
      +\tfrac{1}{2}\varepsilon^{ij}\mathcal{H}_{j}\wedge
      (\star\mathcal{F}-\mathcal{F})
      \nonumber \\
    & \nonumber \\
    & \hspace{.5cm}
      +\tfrac{1}{2}\varepsilon^{ij}
            \left(\mathcal{H}_{j}-\mathcal{M}_{jk}\star\mathcal{H}^{k}\right)\wedge \mathcal{F}
      +\tfrac{1}{2}\frac{\delta G_{A}}{ \delta\mathcal{B}_{i}}\wedge
      \left[\mathfrak{M}^{AB}\star G_{B}-\hat{k}_{B}\right]
      \nonumber \\
    & \nonumber \\
    & \hspace{.5cm}
      +\tfrac{1}{2}\varepsilon^{ij}\mathcal{B}_{j}\wedge \mathbf{E}
      -\tfrac{2}{3}\varepsilon^{ij}\varepsilon^{kl}\mathcal{B}_{j}\wedge
      \mathcal{B}_{k}\wedge \mathbf{E}_{l}\,,
    \\
    & \nonumber \\
        \mathbf{E}
    & =
      -\tfrac{1}{2}
      \left\{ d\star\mathcal{F}
      -\tfrac{1}{2}\varepsilon^{ij} \mathcal{H}_{i}\wedge \left(
      \mathcal{H}_{j}-2\mathcal{M}_{jk}\star \mathcal{H}^{k}\right)\right\}
         +\tfrac{1}{2}\frac{\delta G_{A}}{ \delta\mathcal{D}}\wedge
      \left[\mathfrak{M}^{AB}\star G_{B}-\hat{k}_{B}\right]
      \nonumber \\
    & \nonumber \\
    & \hspace{.5cm}
      +\varepsilon^{ij}\mathcal{B}_{i}\wedge \mathbf{E}_{j}\,,
    \\
    & \nonumber \\
        \mathbf{E}_{i}
    & =
      -\tfrac{1}{2}d\left(\mathcal{M}_{ij}\star\mathcal{H}^{j}\right)
    +\tfrac{1}{2}\frac{\delta G_{A}}{ \delta\mathcal{B}^{i}}\wedge
      \left[\mathfrak{M}^{AB}\star G_{B}-\hat{k}_{B}\right]\,,
    \\
    & \nonumber \\
        \mathbf{E}^{A}
    & =
      -\tfrac{1}{2}d \left[\mathfrak{M}^{AB}\star G_{B}-\hat{k}_{B}\right]\,.
  \end{align}
\end{subequations}

Using the duality relations and the results obtained in the previous
sections we recover the original equations of motion of $\mathcal{N}=2B,d=10$
supergravity.

%%%%%%%%%%%%%%%%%%%%%%%%%%%%%%%%%%%%%%%%%%%%%%%%%%%%%%%%%%%%%%%%%%%%%%
%%%%%%%%%%%%%%%%%%%%%%%%%%%%%%%%%%%%%%%%%%%%%%%%%%%%%%%%%%%%%%%%%%%%%%
%%%%%%%%%%%%%%%%%%%%%%%%%%%%%%%%%%%%%%%%%%%%%%%%%%%%%%%%%%%%%%%%%%%%%%
%%%%%%%%%%%%%%%%%%%%%%%%%%%%%%%%%%%%%%%%%%%%%%%%%%%%%%%%%%%%%%%%%%%%%%
\section{Conclusions}
\label{sec-discussion}
%%%%%%%%%%%%%%%%%%%%%%%%%%%%%%%%%%%%%%%%%%%%%%%%%%%%%%%%%%%%%%%%%%%%%% 
%%%%%%%%%%%%%%%%%%%%%%%%%%%%%%%%%%%%%%%%%%%%%%%%%%%%%%%%%%%%%%%%%%%%%%
%%%%%%%%%%%%%%%%%%%%%%%%%%%%%%%%%%%%%%%%%%%%%%%%%%%%%%%%%%%%%%%%%%%%%%
%%%%%%%%%%%%%%%%%%%%%%%%%%%%%%%%%%%%%%%%%%%%%%%%%%%%%%%%%%%%%%%%%%%%%%

The results obtained in this paper and, in particular, the democratic and
manifestly duality-invariant pseudoactions of $d=4$ maximal and half-maximal
supergravities and of $\mathcal{N}=2B,d=10$ supergravity can be used
in different ways. For instance

\begin{enumerate}
\item They can be used to revisit many of the results on flux
  compactifications and gauged supergravities in a duality-invariant form
  \cite{Cassani:2008rb,Fernandez-Melgarejo:2011nso}. 
\item They can be used to study black-hole thermodynamics using Euclidean
  methods \cite{Gibbons:1976ue}.\footnote{See also \cite{Kallosh:1992wa} and
    references therein.}
\item They can be used to improve our understanding of the interplay between
  T~ and S~dualities. In particular, one can improve our understanding of the
  duality between type~IIB 7- and type~IIA 8-branes
  \cite{Bergshoeff:1996ui,Meessen:1998qm,Fernandez-Melgarejo:2011nso}.
\end{enumerate}

However, several extensions of the results presented here are still necessary:

\begin{enumerate}
\item The supersymmetry transformation rules of all the dual fields we have
  introduced should be found.
  
\item The 10-forms which are known to exist and play an important role in
  $\mathcal{N}=2B,d=10$ supergravity \cite{Bergshoeff:2010mv} should be added
  somehow to the pseudoaction in order to have a complete picture of the
  dualities between fields and fluxes.
  
\end{enumerate}

Work on some of these directions is in progress.

%%%%%%%%%%%%%%%%%%%%%%%%%%%%%%%%%%%%%%%%%%%%%%%%%%%%%%%%%%%%%%%%%%%%%%
%%%%%%%%%%%%%%%%%%%%%%%%%%%%%%%%%%%%%%%%%%%%%%%%%%%%%%%%%%%%%%%%%%%%%%
%%%%%%%%%%%%%%%%%%%%%%%%%%%%%%%%%%%%%%%%%%%%%%%%%%%%%%%%%%%%%%%%%%%%%%
%%%%%%%%%%%%%%%%%%%%%%%%%%%%%%%%%%%%%%%%%%%%%%%%%%%%%%%%%%%%%%%%%%%%%%
\section*{Acknowledgments}
%%%%%%%%%%%%%%%%%%%%%%%%%%%%%%%%%%%%%%%%%%%%%%%%%%%%%%%%%%%%%%%%%%%%%%
%%%%%%%%%%%%%%%%%%%%%%%%%%%%%%%%%%%%%%%%%%%%%%%%%%%%%%%%%%%%%%%%%%%%%%
%%%%%%%%%%%%%%%%%%%%%%%%%%%%%%%%%%%%%%%%%%%%%%%%%%%%%%%%%%%%%%%%%%%%%%
%%%%%%%%%%%%%%%%%%%%%%%%%%%%%%%%%%%%%%%%%%%%%%%%%%%%%%%%%%%%%%%%%%%%%%

The work of JJF-M and GG has been supported in part by the MCI, AEI, FEDER
(UE) grant PID2021-125700NAC22. The work of CG-F, TO and MZ has been supported
in part by the MCI, AEI, FEDER (UE) grants PID2021-125700NB-C21 (``Gravity,
Supergravity and Superstrings'' (GRASS)) and IFT Centro de Excelencia Severo
Ochoa CEX2020-001007-S. The work of GG has been supported by the predoctoral
fellowship FPI-UM R-1006-2021-01. The work of CG-F has been supported by the
MU grant FPU21/02222. The work of MZ has been supported by the fellowship
LCF/BQ/DI20/11780035 from ``La Caixa'' Foundation (ID 100010434). TO wishes to
thank M.M.~Fern\'andez for her permanent support.

%%%%%%%%%%%%%%%%%%%%%%%%%%%%%%%%%%%%%%%%%%%%%%%%%%%%%%%%%%%%%%%%%%%%%% 
%%%%%%%%%%%%%%%%%%%%%%%%%%%%%%%%%%%%%%%%%%%%%%%%%%%%%%%%%%%%%%%%%%%%%%
%%%%%%%%%%%%%%%%%%%%%%%%%%%%%%%%%%%%%%%%%%%%%%%%%%%%%%%%%%%%%%%%%%%%%%
%%%%%%%%%%%%%%%%%%%%%%%%%%%%%%%%%%%%%%%%%%%%%%%%%%%%%%%%%%%%%%%%%%%%%%
%%%%%%%%%%%%%%%%%%%%%%%%%%%%%%%%%%%%%%%%%%%%%%%%%%%%%%%%%%%%%%%%%%%%%%
%%%%%%%%%%%%%%%%%%%%%%%%%%%%%%%%%%%%%%%%%%%%%%%%%%%%%%%%%%%%%%%%%%%%%%
\appendix
%%%%%%%%%%%%%%%%%%%%%%%%%%%%%%%%%%%%%%%%%%%%%%%%%%%%%%%%%%%%%%%%%%%%%%
%%%%%%%%%%%%%%%%%%%%%%%%%%%%%%%%%%%%%%%%%%%%%%%%%%%%%%%%%%%%%%%%%%%%%%
%%%%%%%%%%%%%%%%%%%%%%%%%%%%%%%%%%%%%%%%%%%%%%%%%%%%%%%%%%%%%%%%%%%%%%
%%%%%%%%%%%%%%%%%%%%%%%%%%%%%%%%%%%%%%%%%%%%%%%%%%%%%%%%%%%%%%%%%%%%%%

%%%%%%%%%%%%%%%%%%%%%%%%%%%%%%%%%%%%%%%%%%%%%%%%%%%%%%%%%%%%%%%%%%%%%%
%%%%%%%%%%%%%%%%%%%%%%%%%%%%%%%%%%%%%%%%%%%%%%%%%%%%%%%%%%%%%%%%%%%%%%
%%%%%%%%%%%%%%%%%%%%%%%%%%%%%%%%%%%%%%%%%%%%%%%%%%%%%%%%%%%%%%%%%%%%%%
%%%%%%%%%%%%%%%%%%%%%%%%%%%%%%%%%%%%%%%%%%%%%%%%%%%%%%%%%%%%%%%%%%%%%%

\end{document}